\documentclass[draftclsnofoot,onecolumn,letterpaper,12pt]{IEEEtran}
\usepackage{multirow}
\usepackage{booktabs}
\usepackage{amsfonts}
\usepackage{amsmath,cases}
\usepackage{amssymb}
\usepackage{graphicx}
\usepackage{cite}
\usepackage{epsfig}
\usepackage{url}
\usepackage{algorithm}
\usepackage{algorithmic}
\usepackage{epstopdf}
\usepackage{color}
\usepackage[tight]{subfigure}

\newcommand{\ds}{\displaystyle}
\newtheorem{myth}{Theorem}
\newtheorem{mypro}{Proposition}
\newtheorem{mylem}{Lemma}

\newcommand{\la}{\langle}
\newcommand{\ra}{\rangle}

\newcommand{\qed}{\mbox{}\hfill$\Box$
\vskip 3mm}
\newcommand{\Prf}{{\it Proof: \ }}

\newcommand{\pw}{\pmb{p}}
\newcommand{\bbE}{\mathbb{E}}
\newcommand{\Prob}{\mbox{Prob}}

\newcommand{\br}{\mathbf{r}}
\newcommand{\bp}{\mathbf{p}}
\newcommand{\pk}{p^{(\kappa)}}
\newcommand{\pik}{p_i^{(\kappa)}}
\newcommand{\pjk}{p_j^{(\kappa)}}
\newcommand{\rik}{r_i^{(\kappa)}}
\newcommand{\Rik}{R_i^{(\kappa)}}
\newcommand{\bR}{\mathbf{R}}
\newcommand{\xik}{x_i^{(\kappa)}}

\newcommand{\yijk}{y_i^{(\kappa)}}
\newcommand{\xeik}{x_{e,i}^{(\kappa)}}

\begin{document}
\title{Power Allocation for Energy Efficiency and Secrecy  of Interference  Wireless Networks\thanks{This work was supported in part  by the Australian Research Council’s Discovery Projects under Project DP130104617,  in part by King Fahd University of Petroleum and Minerals under Start-up Research Project \#SR161003,
in part by the U.K. Royal Academy of Engineering Research Fellowship under Grant RF1415$\slash$14$\slash$22, and in part by
the U.S. National Science Foundation under Grants CNS-1702808 and ECCS-1647198 }}
\author{Z. Sheng, H. D. Tuan, A. A. Nasir, T. Q. Duong and H. V.  Poor
\thanks{Zhichao Sheng and Hoang Duong Tuan are with the Faculty of Engineering and Information Technology, University of Technology Sydney, Broadway, NSW 2007, Australia (email: kebon22@163.com, Tuan.Hoang@uts.edu.au)}
\thanks{Ali Arshad Nasir is with Department of Electrical Engineering, King Fahd University of Petroleum and Minerals (KFUPM), Dhahran, Saudi Arabia (email: anasir@kfupm.edu.sa)}
\thanks{Trung Q. Duong is with Queen's University Belfast, Belfast BT7 1NN, UK  (email: trung.q.duong@qub.ac.uk)}
\thanks{H. Vincent Poor is with the Department of Electrical Engineering, Princeton University, Princeton, NJ 08544, USA (e-mail: poor@princeton.edu)}
}
\date{}
\maketitle
\vspace*{-1.8cm}
\begin{abstract}
Considering a multi-user interference network with an eavesdropper, this paper aims at the power allocation
to optimize the worst secrecy throughput among the network links
or the secure energy efficiency in terms of achieved secrecy throughput per Joule under link security requirements.
Three scenarios for the access  of channel state information are considered: the perfect channel state information,
partial channel state information with channels from the transmitters
to the eavesdropper  exponentially distributed, and not perfectly known channels between the transmitters
and the users with exponentially distributed errors.
The paper develops various path-following  procedures of low complexity and rapid convergence
for the optimal power allocation.  Their effectiveness and viability are illustrated through numerical examples.
The power allocation schemes are shown to achieve both high secrecy throughput and energy efficiency.
\end{abstract}
\vspace*{-0.5cm}


\section{INTRODUCTION} \label{sec:Intro}
The broadcast nature of the wireless medium exhibits different challenges in ensuring secure communications in the presence of adversarial users \cite{FTA13,MFHS14}. In particular, it is difficult to protect the transmitted signals from unintended recipients, who may improperly extract information from an ongoing transmission without being detected \cite{LPS08,Baetal13}. Physical layer security \cite{P12,PS17}  has been proposed as a solution to provide security in wireless networks and researchers with a goal being to optimize the secure throughput of a wireless network in the presence of eavesdroppers, which is the difference between the desired user throughput and eavesdroppers' throughput \cite{MFHS14}. Beyond secure throughput, significant interest has recently been put on optimizing the secure energy efficiency (SEE), which is the ratio of the secure throughput to the total network power consumption, measured in terms of bits per Joule per Hertz \cite{WBCH16,NTDP17}.

There has been considerable recent research on physical layer security in wireless communication systems. For example,
assuming the availability of full channel state information (CSI), secrecy optimization has been studied for cooperative relaying networks in \cite{DHPP10,LPW11,Hoetal15}. Energy efficiency (EE) of wireless networks has also drawn attention.  For examples, resource allocation algorithms for the optimization of spectral efficiency as well as EE have been established in \cite{LSYW13}. Keeping EE maximization as an objective, the authors in \cite{RRL17} proposed a precoder design for multi-input-multi-output (MIMO) two-way relay networks. EE maximization for cooperative spectrum sensing in cognitive sensor networks is studied in \cite{ZCLYW17}.

The critical topic of SEE has also been explored very recently \cite{WBCH15,WBCH16,FBSR15,CWMJ16,XYCJ16,OLZZM17,VKT16,NTDP17}.
Specifically, power control algorithms for SEE maximization in decode-and-forward
(DF) and amplify-and-forward (AF) relaying networks have been considered in \cite{WBCH15} and \cite{WBCH16}, respectively. In \cite{FBSR15}, the authors developed a distributed power control algorithm for SEE maximization in DF relaying. The same resource allocation problem for SEE maximization assuming full-duplex relaying is considered in \cite{CWMJ16}.  Recently, the authors in \cite{XYCJ16} and \cite{OLZZM17} also derived the trade-off between SEE and secure spectral efficiency in cognitive radio networks. All these works have assumed the perfect CSI knowledge at the transmitter end, which is not always possible.

It is commonly known that time or frequency resources are generally limited in wireless networks and thus have to be shared among multiple users. This can result in interference among users in the network and thus one has to opt for careful resource allocation or interference alignment schemes \cite{ZKMT14}. Considering a multiuser MIMO interference network,
\cite{VKT16} used the costly interference alignment technique to cancel both information leakage and interference and
then Dinkelbach's method of fractional programming is adopted to optimize EE. As shown in \cite{NTDP17}, both zero-forcing
and interference alignment are not efficient in optimizing the network SEE.

In this paper, we propose novel and efficient resource allocation algorithms for both worst-case secure throughput and worst secure energy efficiency maximization of a highly interference-limited multi-user wireless network. Unlike
many previous works, we do not assume perfect CSI knowledge at the transmitters. In fact, our transmitters only carry channel distribution knowledge for the eavesdropper and imperfect CSI for the users. Particularly, we consider three optimization scenarios to gradually build our algorithms. We start with the ``perfect CSI" scenario. Next, we consider a
``partial CSI" setup where the channel between the transmitters and the eavesdropper is exponentially distributed and only that channel distribution knowledge is available at the transmitters. Finally, we solve for the hardest ``robust optimization" scenario, where in addition to the assumption of only channel distribution knowledge about eavesdroppers, we also assume uncertain channels between the transmitters
and the users with exponentially distributed errors.
We develop various path-following  procedures of low complexity and rapid convergence for the optimal power allocation.  Our extensive simulation results  illustrate their effectiveness and viability.

The rest of the paper is organized as follows. Section II, section III and section IV are devoted to
optimizing the links' worst secrecy throughput and the network secure energy efficiency under the perfect
CSI, partial CSI and imperfectly known CSI, respectively. The simulation is provided in Section V to
show the efficiency of the theoretical developments in the previous section. Appendices provide fundamental rate outage
inequalities and approximations, which are the mathematical base of the theoretical sections II-IV.

\section{Interference networks under perfect CSI}
\begin{figure}
\centering
\includegraphics[width=3.5in]{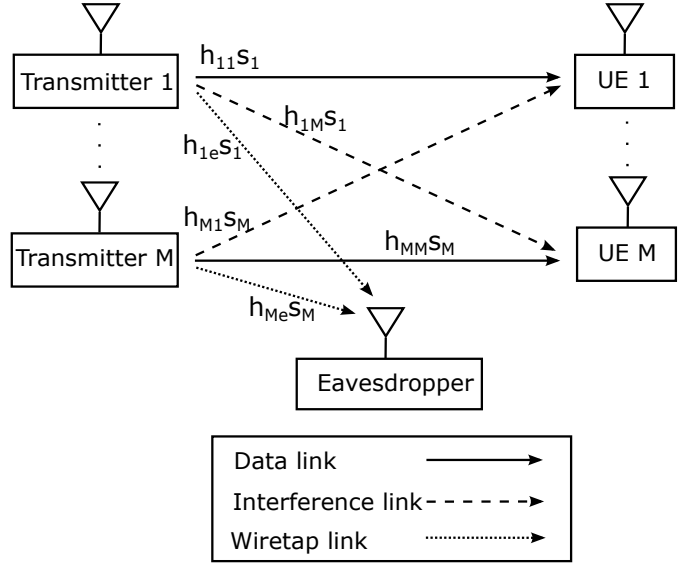}
\caption{System model}
\label{fig:Secrecy_model}
\end{figure}
We consider a cooperative network consisting of $M$ single-antenna transmitters and $M$ single-antenna users
as depicted in Figure \ref{fig:Secrecy_model}, where each transmitter $i$ intends
to send  the information  $s_i$ to user $i$. The information $s_i$
is normalized, i.e. $\mathbb{E}(x_i^2)=1$. Let $\mathbf{p}_i$ be the transmit power allocated to transmitter $i$ and
$\mathbf{p}=(\mathbf{p}_1,\dots,\mathbf{p}_M)^T$. Furthermore, denote by
$h_{ji}$ the channel gain from transmitter $j$ to user $i$. The received signal at user $i$ is
\[
y_i=h_{ii}\bp_i s_i+\sum_{j\neq i}^Mh_{ji}\bp_j s_j+ n_i,
\]
where $n_i\in {\cal CN}(0,\sigma_i^2)$ is additive noise.

Suppose that there is an eavesdropper (EV), which is also equipped with a single antenna. Denoting by
$h_{ie}$ the channel gain from transmitter $i$ to the EV, the received signal at the EV is
\[
y_e=\sum_{i=1}^Mh_{ie}\bp_i s_i+ n_e,
\]
where $n_e\in {\cal CN}(0,\sigma_e^2)$ is additive noise.

Under the perfect CSI at the transmitters, the information throughput
at user $i$ is
\begin{equation}\label{sec1}
f_i(\bp)\triangleq \ln\left(1+\frac{h_{ii}\bp_i}{\sum_{j\neq i}^Mh_{ji}\bp_j+\sigma_i^2}\right).
\end{equation}
With the EV considered as  part of the legitimate network, the channel gain
$h_{ie}$ can also be assumed known \cite{Cetal13}. The wiretapped throughput for
user $i$ at the EV is
\begin{equation}\label{dwire1}
g_i(\bp)\triangleq \ln\left(1+\frac{h_{ie}\bp_i}{\sum_{j\neq i}
h_{je}\bp_j+\sigma_e^2}\right).
\end{equation}
The secrecy throughput in transmitting information $s_i$ to user $i$ while keeping it confidential from the eavesdropper
is defined as
\begin{equation}\label{sec3}
\max \{f_i(\bp)-g_i(\bp),0\}.
\end{equation}
We consider the following fundamental optimization problems in a such network: the maximin secrecy throughput optimization
\begin{subequations}\label{sec4}
\begin{eqnarray}
\max_{\pmb{p}}\ \Phi_{\sf sp}(\bp)\triangleq \min_{i=1,...,M} [ f_i(\bp)-g_i(\bp) ]\quad\mbox{s.t.}\label{sec4a}\\
{\color{black}0<}\bp_i\leq P_i, i=1,\cdots,M,\label{sec4b}
\end{eqnarray}
\end{subequations}
and the network secure energy efficiency (SEE) maximization under users' secrecy throughput quality-of-service (QoS) requirements
\begin{subequations}\label{esec4}
\begin{eqnarray}
\max_{\pmb{p}}\ \Phi_{\sf ee}(\bp)\triangleq\frac{\ds\sum_{i=1}^M[f_i(\bp)-g_i(\bp)]}{\zeta\ds\sum_{i=1}^M\pw_i+P_c} \quad\mbox{s.t.}\quad (\ref{sec4b}),\label{esec4a}\\
f_i(\bp)-g_i(\bp)\geq c_i,\ i=1,..,M,
\label{esec4b}
\end{eqnarray}
\end{subequations}
or the maximin transmitter EE optimization under users' secrecy throughput QoS requirements
\begin{equation}\label{mesec5}
\max_{\pmb{p}}\min_{i=1,...,M} \frac{f_i(\bp)-g_i(\bp)}{{\color{black}\zeta \pw_i}+P_c^i}\quad\mbox{s.t.}\quad (\ref{sec4b}), (\ref{esec4b}).
\end{equation}
Here $\zeta$ is the reciprocal of the drain efficiency of the power amplifier, $P_c^i$
is the circuit power at transmitter $i$ and $P_c=\sum_{i=1}^MP_c^i$. As the numerator
in the objective function in (\ref{esec4}) is the sum secrecy throughput while  the denominator  is the network power consumption, the objective function in (\ref{esec4}) expresses the network SEE in terms of nats/s/Joule. Similarly, each subfunction in  (\ref{mesec5}) expresses the SEE in
for transmitting the information $s_i$. Moreover,
the constraint (\ref{esec4b}) for given thresholds $c_i$ sets the QoS for the users in terms of
the secrecy throughput. This constraint  is nonconvex, which is in contrast to the throughput constraint
\[
f_i(\bp)\geq \tilde{c}_i, i=1,\dots, M,
\]
which is equivalent to the linear constraint
\[
h_{ii}\bp_i\geq (e^{\tilde{c}_i}-1)(\sum_{j\neq i}h_{ji}\bp_j+\sigma_i^2), i=1,\dots, M.
\]
A popular now approach \cite{KTN11} is to treat $f_i-g_i$ in (\ref{sec4})
as a d.c (\underline{d}ifference of two \underline{c}oncave functions) function \cite{Tuybook}:
$f_i(\bp)-g_i(\bp)=\tilde{f}_i(\bp)-\tilde{g}_i(\bp)$ with
$
\tilde{f}_i(\bp)=\ln(\sum_{j=1}^Mh_{ji}\bp_j+\sigma_i^2)+\ln(\sum_{j\neq i}^Mh_{je}\bp_j+\sigma_e^2)
$
and
$\tilde{g}_i(\bp)=\ln(\sum_{j\neq i}^Mh_{ji}\bp_j+\sigma_i^2)+\ln(\sum_{j=1}^Mh_{je}\bp_j+\sigma_e^2)$
which are concave. Then at each iteration, $\tilde{f}_i$ is linearized while $\tilde{g}_i$ is innerly approximated
by a concave quadratic function for a lower approximation of $\tilde{f}_i-\tilde{g}_i$ \cite{Naetal17,Taetal17}. As a result,
each iteration invokes solution of a simple convex quadratic optimization problem with the logarithmic function optimization
of high computational complexity avoided.

Our next subsections are devoted to efficient computational approach to solving each of (\ref{sec4}), (\ref{esec4}) and
(\ref{esec5}) without d.c. representation.
\vspace*{-0.3cm}
\subsection{Max-min secrecy throughput optimization}
At every $p^{(\kappa)}\in R^M_+$, applying  inequality (\ref{sec5.2}) in the Appendix II
for $x=1/h_{ii}\bp_i, y=\sum_{j\neq i}^Mh_{ji}\bp_j+\sigma_i^2$
and $
\bar{x}=1/h_{ii}p_i^{(\kappa)}, \bar{y}=\sum_{j\neq i}^Mh_{ji}p_j^{(\kappa)}+\sigma_i^2$
yields
\begin{eqnarray}
f_i(\bp)\geq f_i^{(\kappa)}(\pmb{p})\label{dc1}
\end{eqnarray}
for
\begin{eqnarray}
f_i^{(\kappa)}(\pmb{p})&\triangleq&\ds\ln(1+\xik)+\ds\frac{\xik}{1+\xik}\left(2-
\frac{\pik}{\bp_i}-
\frac{\sum_{j\neq i}h_{ji}\bp_j+\sigma_i^2}{\sum_{j\neq i}^Mh_{ji}\pjk+\sigma_i^2}\right).\label{dc3}
\end{eqnarray}
On the other hand, applying inequality (\ref{conv1}) in the Appendix II for
$
x=h_{ie}\bp_i, y=\sum_{j\neq i}^Mh_{je}\bp_j
$
and
$
\bar{x}=h_{ie}p_i^{(\kappa)}, \bar{y}=\sum_{j\neq i}^Mh_{je}p_j^{(\kappa)}
$
yields
\begin{equation}\label{dc2}
g_i(\bp)\leq g_i^{(\kappa)}(\pmb{p}),
\end{equation}
for
\begin{eqnarray}
g_i^{(\kappa)}(\pmb{p})&=&\ln(1+\xeik)+\ds\frac{1}{1+\xeik}\left(\frac{0.5h_{ie}(\bp_i^2/\pik+\pik)}{\sum_{j\neq i}^M
h_{je}\bp_j+\sigma_e^2}  -\xeik\right).
\label{dc4}
\end{eqnarray}
Initialized by a feasible $p^{(0)}$ for the convex constraint (\ref{sec4b}), at the $\kappa$-th iteration we solve the convex optimization problem
\begin{equation}\label{dc5}
\max_{\pmb{p}}\ \Phi_{\sf sp}^{(\kappa)}(\bp)\triangleq \min_{i=1,...,M} [f_i^{(\kappa)}(\pmb{p})-g_i^{(\kappa)}(\pmb{p})]\quad\mbox{s.t.}\quad (\ref{sec4b})
\end{equation}
to generate the next iterative point $p^{(\kappa+1)}$.\\
One can see that $\Phi_{\sf sp}(\bp)\geq F_{\sf sp}^{(\kappa)}(\bp)\ \forall\ \bp$ and
$\Phi_{\sf sp}(p^{(\kappa)})=F_{\sf sp}^{(\kappa)}(p^{(\kappa)})$. Futhermore,
$\Phi_{\sf sp}^{(\kappa)}(p^{(\kappa+1)})>\Phi_{\sf sp}^{(\kappa)}(p^{(\kappa)})$ if
$p^{(\kappa+1)}\neq p^{(\kappa)}$ because the former is the optimal solution of
(\ref{dc5}) while the latter is its feasible point. Therefore,
\begin{equation}\label{pathf}
\Phi_{\sf sp}(p^{(\kappa+1)})\geq  \Phi^{(\kappa)}_{\sf sp}(p^{(\kappa+1)})> \Phi^{(\kappa)}_{\sf sp}(p^{(\kappa)})
=\Phi_{\sf sp}(p^{(\kappa)}),
\end{equation}
i.e. $p^{(\kappa+1)}$ is better than $p^{(\kappa)}$; as such $\{p^{(\kappa)}\}$ is a sequence of improved points
that converges at least to a locally optimal solution of (\ref{sec4}) satisfying the first order necessary optimality condition \cite[Prop. 1]{MW78}. In summary,  we propose in Algorithm \ref{alg1}
a path-following computational procedure for the maximin secrecy throughput optimization problem (\ref{sec4}).
\begin{algorithm}
\caption{Path-following algorithm for maximin secrecy throughput optimization} \label{alg1}
\begin{algorithmic}
\STATE \textbf{Initialization}: Set $\kappa=0$. Choose an initial feasible point
$p^{(0)}$ for the convex constraints (\ref{sec4b}).
Calculate $R_{\min}^{(0)}$ as the value of the objective in (\ref{sec4}) at $p^{(0)}$.
\REPEAT \STATE $\bullet$ Set
$\kappa=\kappa+1$.
\STATE $\bullet$ Solve the
convex optimization problem \eqref{dc5} to obtain the
solution $p^{(\kappa)}$.
\STATE $\bullet$ Calculate $R_{\min}^{(\kappa)}$ as the value of the objective in (\ref{sec4}) at $p^{(\kappa)}$.
 \UNTIL{$\frac{R_{\min}^{(\kappa)}-R_{\min}^{(\kappa-1)})}{R_{\min}^{(\kappa-1)}} \leq
 \epsilon_{\rm tol}$}.
\end{algorithmic}
\end{algorithm}
\vspace*{-0.7cm}
\subsection{Secure energy efficient maximization}
Define
\[
\pi(\bp)=\zeta\ds\sum_{i=1}^M\bp_i+P_c.
\]
Applying the inequality (\ref{esec5.2}) in Appendix II for
$
x=1/h_{ii}\bp_i$, $y=\sum_{j\neq i}^Mh_{ji}\bp_j+\sigma_i^2, t=\pi(\bp)$,
and $\bar{x}=1/h_{ii}p_i^{(\kappa)}$, $\bar{y}=\sum_{j\neq i}^Mh_{ji}p_j^{(\kappa)}+\sigma_i^2, \bar{t}=\pi(p^{(\kappa)})$
yields
\begin{eqnarray}\label{fedc3}
\ds\frac{f_i(\bp)}{\pi(\bp)}&\geq&F_i^{(\kappa)}(\bp)
\end{eqnarray}
for
\begin{eqnarray}
F_i^{(\kappa)}(\bp)&\triangleq&  \ds\frac{2\ln(1+\xik)}{\pi(\pk)}
+\ds\frac{\xik}{\pi(\pk)(1+\xik)}\left(2-
\frac{\pik}{\bp_i}-
\frac{\sum_{j\neq i}h_{ji}\bp_j+\sigma_i^2}{\sum_{j\neq i}^Mh_{ji}\pjk+\sigma_i^2}\right)\nonumber\\
&&-\ds\frac{\ln(1+\xik)}{\pi^2(\pk)}\pi(\bp).\label{edc3}
\end{eqnarray}
On the other hand, applying inequality (\ref{conv1}) in Appendix II for
$
\alpha=1+\ln(2)$, $x=h_{ie}\bp_i/(\sum_{j\neq i}h_{je}\bp_j+\sigma_e^2), t=\pi(\bp)
$
and
$
\bar{x}\triangleq h_{ie}\pik/(\sum_{j\neq i}
h_{je}\pjk+\sigma_e^2)$, $\bar{t}=\pi(p^{(\kappa)})$
yields
\begin{eqnarray}
\ds\frac{-g_i(\bp)}{\pi(\bp)}&\geq&\ds\ 2\frac{\alpha-\ln(1+\xeik)}{\pi(\pk)}+\frac{\xeik}{(1+\xeik)\pi(\pk)}
-\ds\frac{1}{(1+\xeik)\pi(\pk)}\frac{h_{ie}\bp_i}{\sum_{j\neq i}
h_{je}\bp_j+\sigma_e^2}\nonumber\\
&&-\frac{\alpha-\ln(1+\xeik)}{\pi^2(\pk)}\pi(\bp)-\frac{\alpha}{\pi(\bp)}
\end{eqnarray}
which together with (\ref{conv3}) in Appendix II yield
\begin{equation}\label{nin2}
\ds\frac{f_i(\bp)-g_i(\bp)}{\pi(\bp)}\geq G_i^{(\kappa)}(\bp)
\end{equation}
for the concave function
\begin{eqnarray}
G_i^{(\kappa)}(\bp)&\triangleq&\ds\ 2\frac{\alpha-\ln(1+\xeik)}{\pi(\pk)}+\frac{\xeik}{(1+\xeik)\pi(\pk)}
-\ds\frac{1}{(1+\xeik)\pi(\pk)}\frac{0.5h_{ie}(\bp_i^2/\pik+\pik)}{\sum_{j\neq i}
h_{je}\bp_j+\sigma_e^2}\nonumber\\
&&-\ds\frac{\alpha-\ln(1+\xeik)}{\pi^2(\pk)}\pi(\bp)-\frac{\alpha}{\pi(\bp)}.\label{nin3}
\end{eqnarray}
Initialized by a feasible point $p^{(0)}$ for (\ref{esec4}),
we solve the following convex optimization problem at the $\kappa$-th iteration to generate
the next iterative point $p^{(\kappa+1)}$:
\begin{subequations}\label{nin5}
\begin{eqnarray}
\max_{\pmb{p}}\ \Phi_{\sf ee}^{(\kappa)}(\bp)\triangleq  \sum_{i=1}^M[F_i^{(\kappa)}(\pmb{p})+G_i^{(\kappa)}(\pmb{p})]\quad\mbox{s.t.}\quad
(\ref{sec4b}), \label{nin5a}\\
f_i^{(\kappa)}(\pmb{p})-g_i^{(\kappa)}(\pmb{p})\geq c_i, i=1,\dots,M.\label{nin5b}
\end{eqnarray}
\end{subequations}
Due to (\ref{dc1}) and (\ref{dc2}), the nonconvex constraint (\ref{esec4b}) in (\ref{esec4}) is implied by
the convex constraint (\ref{nin5b}) in (\ref{nin5}). Similarly to (\ref{pathf}), we can show that $\Phi_{\sf ee}(p^{(\kappa+1)})>\Phi_{\sf ee}(p^{(\kappa)})$ whenever
$p^{(\kappa+1)}\neq p^{(\kappa)}$; as such the computational procedure that invokes the convex program
(\ref{nin5}) to generate the next iterative point, is path-following for (\ref{esec4}), which at least converges to its locally optimal solution satisfying the Karush-Kuh-Tucker (KKT) conditions of optimality.

\noindent Recalling the definition (\ref{dc2}) and (\ref{dc4}) of functions $f_i^{(\kappa)}$ and $g_i^{(\kappa)}$, initialized by any feasible point $\tilde{p}^{0)}$ for
the convex constraint (\ref{sec4b}), we generate $\tilde{p}^{(\kappa+1)}$, $\kappa=0, \dots$, as the optimal solution of the convex optimization problem
\begin{equation}\label{feap}
\max_{\bp}\ \min_{i=1,\dots,M} \frac{f_i^{({\kappa})}(\bp)-g_i^{(\kappa)}(\bp)}{c_i}\quad
\mbox{s.t.}\quad (\ref{sec4b})
\end{equation}
until $\tilde{p}^{(\kappa+1})$ such that
$\min_{i=1,\dots,M} (f_i(p^{(\kappa+1)})-g_i(p^{(\kappa+1))}/c_i
\geq 1$ is found and thus $p^{(0)}=\tilde{p}^{(\kappa+1})$ is feasible for (\ref{esec4}) that is needed for
the initial step.

Analogously, to address the maximin secure energy efficient optimization problem (\ref{esec5}) define
\[
\pi_i(\bp_i)=\zeta \bp_i +P_c^i.
\]
Similarly to (\ref{fedc3}) and (\ref{nin2}) the following inequalities can be obtained:
\begin{eqnarray}
\ds\frac{f_i(\bp)}{\pi_i(\pw_i)}&\geq&\tilde{F}_i^{(\kappa)}(\bp_i)
\label{ext1}\\
\ds\frac{-g_i(\bp)}{\pi_i(\bp_i)}&\geq& \tilde{G}_i^{(\kappa)}(\bp_i)\label{ext2}
\end{eqnarray}
for
\begin{eqnarray}
\tilde{F}_i^{(\kappa)}(\bp_i)&\triangleq&  \ds\frac{2\ln(1+\xik)}{\pi_i(\pik)}+\ds\frac{\xik}{\pi_i(\pik)(1+\xik)}\left(2-
\frac{\pik}{\bp_i}-
\frac{\sum_{j\neq i}h_{ji}\bp_j+\sigma_i^2}{\sum_{j\neq i}^Mh_{ji}\pjk+\sigma_i^2}\right)\nonumber\\
&&-\ds\frac{\ln(1+\xik)}{\pi_i^2(\pik)}\pi_i(\bp_i)\label{ext3}\\
\tilde{G}_i^{(\kappa)}(\bp_i)&\triangleq& \ds\ 2\frac{\alpha-\ln(1+\xeik)}{\pi_i(\pik)}+\frac{\xeik}{(1+\xeik)\pi_i(\pik)}
-\ds\frac{1}{(1+\xeik)\pi_i(\pik)}\frac{0.5h_{ie}(\bp_i^2/\pik+\pik)}{\sum_{j\neq i}
h_{je}\bp_j+\sigma_e^2}\nonumber\\
&&-\ds\frac{\alpha-\ln(1+\xeik)}{\pi_i^2(\pik)}\pi_i(\bp_i)-\frac{\alpha}{\pi_i(\bp_i)}.\label{ext4}
\end{eqnarray}
Initialized  a feasible point $p^{(0)}$ for (\ref{esec5}), which is found by using the generation (\ref{feap}),
the following convex optimization problem at the $\kappa$-th iteration is proposed to generate
the next iterative point $p^{(\kappa+1)}$:
\begin{equation}\label{mmee}
\max_{\pmb{p}}\min_{i=1,...,M}  [\tilde{F}_i^{(\kappa)}(\pmb{p})+\tilde{G}_i^{(\kappa)}(\pmb{p})]\quad\mbox{s.t.}\quad
(\ref{sec4b}), f_i^{(\kappa)}(\pmb{p})-g_i^{(\kappa)}(\pmb{p})\geq c_i, i=1,\dots,M.
\end{equation}
The computational procedure that invokes the convex program
(\ref{mmee}) to generate the next iterative point, is path-following for (\ref{esec5}), which at least converges to its locally optimal solution satisfying the first order necessary optimality condition.
\section{Interference networks under partial wiretap CSI}
When the EV is not part of the legitimate network, it is almost impossible to estimate
channels $h_{ie}$ from the transmitters to it. It is common to  assume
that $h_{ie}\sim \bar{h}_{ie}\chi_{ie}$, where $\chi_{ie}$ is an exponential distribution with the unit mean
and $\bar{h}_{ie}$ is a known deterministic quantity. Accordingly, instead of (\ref{dwire1}),
the wiretapped throughput for user $i$ at the
EV is defined via the following throughput outage  \cite{KB02,GRZ10,LCLC13,LCC15a,LCC15b}:
\begin{equation}\label{wire1}
g_{i,o}(\bp)\triangleq \max\ \{ \ln(1+\mathbf{r}_i)\ :\ \Prob\left(\frac{h_{ie}\bp_i}{\sum_{j\neq i}
h_{je}\bp_j+\sigma_e^2}<\mathbf{r}_i\right)\leq \epsilon_{EV} \}
\end{equation}
for  $\epsilon_{EV}>0$. Using (\ref{root1}) in Appendix I, it follows that
\[
g_{i,o}(\bp)=\ln(1+\mathbf{r}_i)
\]
where
\begin{equation}\label{esec5b}
\bp_i\bar{h}_{ie}\ln(1-\epsilon_{EV})+\ds\br_{i}\sigma_e^2
\bp_i\bar{h}_{ie}\sum_{j\neq i}^M\ln\left(1+\frac{\br_{i}\bar{h}_{je}\bp_j}{\bar{h}_{ie}\bp_i}\right)= 0,
\ i=1,...,M.
\end{equation}
Therefore, the problem of maximin secrecy throughput optimization can be formulated as
\begin{subequations}\label{esec5}
\begin{eqnarray}
\max_{\pmb{p},\br}\min_{i=1,...,M} [f_i(\bp)-\ln(1+\br_i)]\quad\mbox{s.t}\quad (\ref{sec4b}), (\ref{esec5b}),\label{esec5a}\\
{\color{black}\br_i>0, i=1,...,M.}\label{sec5c}
\end{eqnarray}
\end{subequations}
The following result unravels the computationally intractable nonlinear equality constraints in (\ref{esec5b}).
\begin{mypro}\label{outpro} The problem (\ref{esec5}) is equivalent to the following problem
\begin{subequations}\label{sec5}
\begin{eqnarray}
\max_{\pmb{p},\br}\min_{i=1,...,M} [f_i(\bp)-\ln(1+\br_i)]\quad\mbox{s.t}\quad (\ref{sec4b}), (\ref{sec5c}),\label{sec5a}\\
\bp_i\bar{h}_{ie}\ln(1-\epsilon_{EV})+\ds\br_{i}\sigma_e^2+
\bp_i\bar{h}_{ie}\sum_{j\neq i}^M\ln\left(1+\frac{\br_{i}\bar{h}_{je}\bp_j}{\bar{h}_{ie}\bp_i}\right){\color{black}\geq} 0,
\ i=1,...,M.
\label{sec5b}
\end{eqnarray}
\end{subequations}
\end{mypro}
\Prf Since the equality constraint (\ref{esec5b}) implies the inequality constraint (\ref{sec5b}), it is true that
\[
\max\ (\ref{esec5}) \leq \max\ (\ref{sec5}).
\]
We now show that there is an optimal solution of (\ref{sec5}) satisfies the equality constraint (\ref{esec5b}) and thus
\[
\max\ (\ref{sec5}) \leq \max\ (\ref{esec5}),
\]
showing the equivalence between (\ref{sec5}) and (\ref{esec5}). Indeed, suppose that at the optimality,
\[
\bp_i\bar{h}_{ie}\ln(1-\epsilon_{EV})+\ds\br_{i}\sigma_e^2+
\bp_i\bar{h}_{ie}\sum_{j\neq i}^M\ln\left(1+\frac{\br_{i}\bar{h}_{je}\bp_j}{\bar{h}_{ie}\bp_i}\right)> 0
\]
for some $i=1, \dots, M$. Then there is $0<\gamma_i<1$ such that
\[
\bp_i\bar{h}_{ie}\ln(1-\epsilon_{EV})+\ds(\gamma\br_{i})\sigma_e^2+
\bp_i\bar{h}_{ie}\sum_{j\neq i}^M\ln\left(1+\frac{\gamma\br_{i}\bar{h}_{je}\bp_j}{\bar{h}_{ie}\bp_i}\right)=0,
\]
that yields
\[
f_i(\bp)-\ln(1+\gamma\br_i)> f_i(\bp)-\ln(1+\br_i),
\]
so $\gamma_i\mathbf{r}_i$ is also the optimal solution of (\ref{sec5}), which satisfies the equality constraint
(\ref{esec5b}).\qed

To address problem (\ref{sec5}), note that a lower bounding function for the first term in (\ref{sec5a}) is $f_i^{(\kappa)}(\pmb{p})$ defined by (\ref{dc3}), while an upper bounding function
for the second term in (\ref{sec5a}) is the following linear function
\begin{equation}\label{dc6}
a_i^{(\kappa)}(\br_i)=\ln(1+\rik)-\ds\frac{\rik}{\rik+1}+\frac{\br_i}{\rik+1}.
\end{equation}
The main difficulty now is to develop a lower bounding approximation for the function in the left hand side (LHS)
of constraint (\ref{sec5b}). Applying  inequality (\ref{sec5.2}) in Appendix II
for
$x=1/\br_i\bar{h}_{je}\bp_j$, $y=\bar{h}_{ie}\bp_i$
and
$\bar{x}=1/\rik\bar{h}_{je}\pjk$, $\bar{y}=\bar{h}_{ie}\pik$
yields
\begin{eqnarray}
\ds\ln\left(1+\frac{\br_{i}\bar{h}_{je}\bp_j}{\bar{h}_{ie}\bp_i}\right)&\geq&\lambda_{ij}^{(\kappa)}(\br_i,\pmb{p}_j,\pmb{p}_i)
\label{sec5.3}
\end{eqnarray}
for
\begin{eqnarray}\label{dsec5.3}
\lambda_{ij}^{(\kappa)}(\br_i,\pmb{p}_j,\pmb{p}_i)
&\triangleq&\ds\ln(1+x_{ij}^{(\kappa)})+y_{ij}^{(\kappa)}\left(2-\frac{\rik\pjk}{\br_i\bp_j}-\frac{\bp_i}{\pik}\right)
\end{eqnarray}
with $x_{ij}^{(\kappa)}\triangleq
\rik\bar{h}_{je}\pjk/\bar{h}_{ie}\pik$ and $y_{ij}^{(\kappa)}\triangleq x_{ij}^{(\kappa)}/(x_{ij}^{(\kappa)}+1)$.
Therefore, over the trust region
\begin{equation}\label{tr1}
\begin{array}{c}
\lambda_{ij}^{(\kappa)}(\br_i,\pmb{p}_j,\pmb{p}_i)   \geq 0,\\
2.5-\ds\frac{\br_i}{\rik}-\frac{\bp_j}{\pjk}\geq 0
\end{array}
\end{equation}
it is true that
\begin{eqnarray}
\bp_i\ds\ln(1+\frac{\br_{i}\bar{h}_{je}\bp_j}{\bar{h}_{ie}\bp_i})&\geq&
\bp_i\ds\ln(1+x_{ij}^{(\kappa)})+y_{ij}^{(\kappa)}\left(2\bp_i-\frac{\rik\pjk\bp_i}{\br_i\bp_j}-\frac{\bp_i^2}{\pik}\right)
\nonumber\\
&=&\ds\left(\ln(1+x_{ij}^{(\kappa)})+2y_{ij}^{(\kappa)}\right)\bp_i-0.5y_{ij}^{(\kappa)}\left[2\frac{\bp_i^2}{\pik}
\right.\nonumber\\
&&\left.\ds+(\frac{\sqrt{2}\bp_i}{\sqrt{\pik}}+\frac{\sqrt{\pik}\rik\pjk}{\sqrt{2}\br_i\bp_j})^2-
\frac{2\bp_i^2}{\pik}-\frac{\pik(\rik\pjk)^2}{2\br_i^2\bp_j^2}
\right]\nonumber\\
&\geq&\Lambda_i^{(\kappa)}(\br_i,\bp_j,\bp_i)\label{ob1}
\end{eqnarray}
for
\[
\begin{array}{lll}
\Lambda_i^{(\kappa)}(\br_i,\bp_j,\bp_i)
&=&\ds\left(\ln(1+x_{ij}^{(\kappa)})+2y_{ij}^{(\kappa)}\right)\bp_i-0.5\yijk\left(\frac{\sqrt{2}\bp_i}{\sqrt{\pik}}+
\frac{\sqrt{\pik}\rik\pjk}{\sqrt{2}\br_i\bp_j}\right)^2\\
&&- \ds{\color{black}0.5\yijk}\pik\left(\frac{\br_i}{\rik}
+\frac{\bp_j}{\pjk}-2.5\right).
\end{array}
\]
Note that in obtaining (\ref{ob1})  we also used the fact that function $\nu(\br_i,\bp_j)\triangleq 1/\br_i^2\bp_j^2$ is convex in the domain $\{ \br_i>0, \bp_j>0\}$ and accordingly \cite{Tuybook}
$1/\br_i^2\bp_j^2\geq \nu(\rik,\pjk)+\la \nabla \nu(\rik,\pjk), (\br_i,\bp_j)-(\rik,\pjk)\ra=
[5-2(\br_i/\rik+\bp_j/\pjk)]/(\rik\pjk)^2$.

Initialized from a feasible point $(p^{(0)}, r^{(0)})$ for (\ref{sec5}) we solve the following convex program at the
$\kappa$-th iteration to generate $(p^{(\kappa+1)}, r_u^{(\kappa+1)})$:
\begin{subequations}\label{sec5k}
\begin{eqnarray}
\max_{\pmb{w},\br}\min_{i=1,...,M} [f_i^{(\kappa)}(\pmb{p})-a_i^{(\kappa)}(\br_i)]\quad\mbox{s.t}\quad (\ref{sec4b}), {\color{black}(\ref{sec5c})}, (\ref{tr1}), \label{sec5ka}\\
\pmb{p}_i\bar{h}_{ie}\ln(1-\epsilon_{EV})+\sigma_e^2\br_i+
\bar{h}_{ie}\sum_{j\neq i}^M\Lambda_{ij}^{(\kappa)}(\br_i,\pmb{p}_j,\pmb{p}_i){\color{black}\geq} 0,
\ i=1,...,M.
\label{sec5kb}
\end{eqnarray}
\end{subequations}
Then $r_i^{(\kappa+1)}$ is found from solving
\begin{eqnarray}
\psi_i(\br_i)\triangleq p_i^{(\kappa+1)}\bar{h}_{ie}\ln(1-\epsilon_{EV})+\ds\br_{i}\sigma_e^2 +
p_i^{(\kappa+1)}\bar{h}_{ie}\sum_{j\neq i}^M\ln\left(1+\frac{\br_{i}\bar{h}_{je}p_j^{(\kappa+1)}}{\bar{h}_{ie}p_i^{(\kappa+1)}}\right)= 0,\label{bi1}\\
\ i=1,...,M,\nonumber
\end{eqnarray}
by bisection on $[0,r_{u,i}^{(\kappa+1)}]$ such that
\begin{equation}\label{bi2}
0\leq \psi_i(r_i^{(\kappa+1)})\leq \epsilon_b\quad\mbox{(tolerance)}.
\end{equation}
A bisection on $[r_{l}, r_u]$
for solving $\psi_i(\mathbf{r}_i)=0$ where $\psi_i$ increases in $\mathbf{r_i}>0$ is implemented as
follows:
\begin{itemize}
\item Define $r_i=(r_l+r_u)/2$.  Reset  $r_l=r_i$
if $\psi_i(r_i)<0$. Otherwise reset $r_u=r_i$.
\item Terminate until $0\leq \psi_i(r_i)\leq \epsilon_b$.
\end{itemize}
In summary,  we propose in Algorithm \ref{alg2}
a path-following computational procedure for the maximin secrecy throughput optimization problem (\ref{sec5}),
which at least converges to its locally optimal solution satisfying the first order necessary optimality condition.
\begin{algorithm}
\caption{Path-following algorithm for maximin secrecy throughput optimization } \label{alg2}
\begin{algorithmic}
\STATE \textbf{Initialization}: Set $\kappa=0$. Choose an initial feasible point
$(p^{(0)}, r^{(0)})$ for (\ref{sec5}) and calculate $R_{\min}^{(0)}$ as the value of the objective function in
(\ref{sec5}) at $(p^{(0)}, r^{(0)})$.
\REPEAT \STATE $\bullet$ Set
$\kappa=\kappa+1$.
\STATE $\bullet$ Solve the
convex optimization problem \eqref{sec5k} to obtain the
solution $(p^{(\kappa)}, r_u^{(\kappa)})$.
\STATE $\bullet$ Solve the nonlinear equations (\ref{bi1}) to obtain the roots $r_i^{(\kappa)}$.
\STATE $\bullet$ Calculate  $R_{\min}^{(\kappa)}$ as the value of the objective function in (\ref{sec5}) at
$(p^{(\kappa)}, r^{(\kappa)})$.
 \UNTIL{$\frac{R_{\min}^{(\kappa)}-R_{\min}^{(\kappa-1)})}{R_{\min}^{(\kappa-1)}} \leq
 \epsilon_{\rm tol}$}.
\end{algorithmic}
\end{algorithm}

A feasible $(p^{(0)}, r^{(0)})$ is found as follows: taking $p^{(0)}$ feasible to the power constraint (\ref{sec4b})
and finding $r^{(0)}$ from solving
\[
\psi_i(\br_i)\triangleq p_i^{(0)}\bar{h}_{ie}\ln(1-\epsilon_{EV})+\ds \br_{i}\sigma_e^2 +
\bar{h}_{ie}p_i^{(0)}\sum_{j\neq i}^M\ln\left(1+\frac{\br_{i}\bar{h}_{je}p_j^{(0)}}{\bar{h}_{ie}p_i^{(0)}}\right)= 0,
\ i=1,...,M,
\]
by  bisection on $[0,r_{u,i}^{(0)}]$ with $\psi_i(r_{u,i})>0$. Such $r^{(0)}_{u,i}$ can be easily found: from any $r_{u,i}>0$, if
$\psi_i(r_{u,i})\geq 0$ then we are done. Otherwise reset $r_{u,i}\leftarrow 2r_{u,i}$ and check $\psi_i(r_{u,i})$. Stop when $\psi(r_{u,i})>0$. Intuitively, taking $r_{u,i}^{(0)}=\bar{h}_{ie}p_i^{(0)}/\sigma_e^2$ will work.

Furthermore, the problem of SEE maximization can be formulated as
\begin{subequations}\label{esec55}
\begin{eqnarray}
\ds\max_{\pmb{p},\br}\frac{\sum_{i=1}^M \left(f_i(\bp)-\ln(1+\br_i)\right)}{\pi(\bp)}\quad\mbox{s.t}\quad (\ref{sec4b}), (\ref{sec5c}), (\ref{sec5b}) \label{esec55a}\\
f_i(\bp)-\ln(1+\br_i)\geq c_i, i=1,\dots, M.\label{esec55b}
\end{eqnarray}
\end{subequations}
Using the inequality (\ref{nin1}) in Appendix II leads to
\[
\frac{-\ln(1+\br_i)}{\pi(\bp)}\geq \tilde{a}_i^{(\kappa)}(\br_i,\bp)
\]
for
\begin{eqnarray}
\tilde{a}_i^{(\kappa)}(\br_i,\bp)&\triangleq&2\ds\frac{\alpha-\ln(1+\rik)}{\pi(\pk)}+
\frac{\rik}{\pi(\pk)(1+\rik)}-\frac{\br_i}{\pi(\pk)(1+\rik)}\nonumber\\
&&-\ds\frac{\alpha-\ln(1+\rik)}{\pi^2(\pk)}\pi(\bp)-\frac{\alpha}{\pi(\bp)}.\label{enin1}
\end{eqnarray}
Initialized by a feasible $(p^{(0)},r^{(0)})$, the following convex programm
is solved to generate $(p^{(\kappa+1)}, r^{(\kappa+1)})$ at the $\kappa$iteration:
\begin{subequations}\label{esec5k}
\begin{eqnarray}
\ds\max_{\pmb{p},\br}\sum_{i=1}^M [F_i^{(\kappa)}(\pmb{p})+\tilde{a}_i^{(\kappa)}(\br_i,\bp)]\quad\mbox{s.t}\quad (\ref{sec4b}), {\color{black}(\ref{sec5c})}, (\ref{tr1}), (\ref{sec5kb}),\label{esec5ka}\\
f_i^{(\kappa)}(\pmb{p})-a_i^{(\kappa)}(\br_i)\geq c_i, i=1,\dots, M.
\end{eqnarray}
\end{subequations}
It can be shown that the computational procedure that invokes the convex program (\ref{esec5k}) to generate
the next iterative point, is path-following for (\ref{esec55}), which at least converges to its locally optimal solution satisfying the KKT conditions.\\
A point $(p^{(0)},r^{(0)})$ is feasible for (\ref{esec55}) if and only if
$\min_{i=1,\dots,M}[f_i(p^{(0)})-\ln(1+r_i^{(0)})]/c_i\geq 1$ and thus can be easily located by adapting Algorithm
\ref{alg2}.

Similarly, a path-following procedure for the following maximin SEE optimization problem can be proposed
\begin{equation}\label{msee}
\ds\max_{\pmb{p},\br}\min_{i=1,\dots,M}\frac{f_i(\bp)-\ln(1+\br_i)}{\pi_i(\bp)}\quad\mbox{s.t}\quad (\ref{sec4b}), (\ref{sec5c}), (\ref{sec5b}), (\ref{esec55b}).
\end{equation}
\section{Robust Optimization}
Beside assuming that $h_{ie}\sim \bar{h}_{ie}\chi_{ie}$
with an exponential distribution $\chi_{ie}$ with the unit mean and deterministic $h_{ie}$,
we also assume that CSI of $h_{ji}$ is not known perfectly in the form
$h_{ji}=\bar{h}_{ji}(1+\delta \chi_{ji})$ with deterministic $\bar{h}_{ji}$
and $\delta$, and random  $\chi_{ji}$, which is an independent exponential distribution of the unit mean,
and $h_{ie}\sim \bar{h}_{ie}\chi_{ie}$ with exponential distributions  $\chi_{ji}$ and $\chi_{ie}$ of the unit mean.
Instead of (\ref{sec1}), the throughput at user $i$ is defined via the following outage probability
\begin{equation}\label{rotage1}
f_{i,o}(\bp)\triangleq \max \{ \ln(1+\bR_i)\ :\ \Prob\left(\frac{h_{ii}\bp_i}{\sum_{j\neq i}^Mh_{ji}\bp_j+\sigma_i^2}<\bR_i\right)\leq
\epsilon_c\}
\end{equation}
for $0<\epsilon_c<<1$.\\
Using (\ref{root2}) in Appendix II, it follows that
\begin{equation}\label{rotage2}
f_{i,o}(\bp)=\ln(1+\bR_i), i=1,\dots, M,
\end{equation}
where
\begin{eqnarray}
\pmb{p}_{i}\bar{h}_{ii}[\delta\ln(1-\epsilon_c)-1]+\ds\bR_{i}(\sigma_i^2+\sum_{j\neq i}\bar{h}_{ji}\bp_j) +
\delta\bar{h}_{ii}\pmb{p}_{i}\sum_{j\neq i}^M\ln\left(1+\frac{\bar{h}_{ji}\bR_{i}\pmb{p}_{j}}{\bar{h}_{ii}\pmb{p}_{i}}
\right)=0, \label{rotage3}\\
 i=1,\dots, M.\nonumber
\end{eqnarray}
Therefore, the problem of maximin secrecy throughput robust optimization is defined by
\begin{subequations}\label{rotage4}
\begin{eqnarray}
\max_{\pmb{p},\bR,\br}\min_{i=1,...,M} [\ln(1+\bR_i)-\ln(1+\br_i)]
\quad\mbox{s.t}\quad (\ref{sec4b}),  {\color{black}(\ref{sec5c})}, (\ref{sec5b}), (\ref{rotage3}), \label{rotage4a}\\
{\color{black}\bR_i>0, i=1,...,M.}\label{sec6c}
\end{eqnarray}
\end{subequations}
The following result unravels the computationally intractable nonlinear equality constraints in (\ref{rotage3}):
\begin{mypro}\label{prorob}
Problem (\ref{rotage4}) is equivalent to the following problem
\begin{subequations}\label{sec6}
\begin{eqnarray}
\max_{\pmb{p},\bR,\br}\min_{i=1,...,M} [\ln(1+\bR_i)-\ln(1+\br_i)]
\quad\mbox{s.t}\quad (\ref{sec4b}), (\ref{sec5b}), {\color{black}(\ref{sec5c})}, (\ref{sec6c})\label{sec6a}\\
\pmb{p}_{i}\bar{h}_{ii}[\delta\ln(1-\epsilon_c)-1]+\ds\bR_{i}(\sigma_i^2+\sum_{j\neq i}\bar{h}_{ji}\bp_j) +
\delta\bar{h}_{ii}\pmb{p}_{i}\sum_{j\neq i}^M\ln\left(1+\frac{\bar{h}_{ji}\bR_{i}\pmb{p}_{j}}{\bar{h}_{ii}\pmb{p}_{i}}\right)\leq
0,\label{sec6be}\\
\ i=1,...,M. \nonumber
\end{eqnarray}
\end{subequations}
\end{mypro}
\Prf Again, it is obvious that
\[
\max\ (\ref{rotage4}) \leq (\ref{sec6}).
\]
Furthermore, at an optimal solution of (\ref{sec6}), if
\[
\pmb{p}_{i}\bar{h}_{ii}[\delta\ln(1-\epsilon_c)-1]+\ds\bR_{i}(\sigma_i^2+\sum_{j\neq i}\bar{h}_{ji}\bp_j) +
\delta\bar{h}_{ii}\pmb{p}_{i}\sum_{j\neq i}^M\ln\left(1+\frac{\bar{h}_{ji}\bR_{i}\pmb{p}_{j}}{\bar{h}_{ii}\pmb{p}_{i}}\right)<0,
\]
for some $i$ then there is $\gamma>1$ such that
\[
\pmb{p}_{i}\bar{h}_{ii}[\delta\ln(1-\epsilon_c)-1]+\ds(\gamma\bR_{i})(\sigma_i^2+\sum_{j\neq i}\bar{h}_{ji}\bp_j) +
\delta\bar{h}_{ii}\pmb{p}_{i}\sum_{j\neq i}^M\ln\left(1+\frac{\bar{h}_{ji}(\gamma\bR_{i})\pmb{p}_{j}}{\bar{h}_{ii}\pmb{p}_{i}}\right)=0,
\]
which results in $\ln(1+\gamma \bR_i)>\ln(1+\bR_i)$, implying that $\gamma \bR_i$ is also an optimal solution
of (\ref{sec6}). We thus have proved that there is always an optimal solution of (\ref{sec6}) to satisfy
the equality constraints in (\ref{rotage3}), so
\[
\max\ (\ref{sec6}) \leq (\ref{rotage4}),
\]
completing the proof of Proposition \ref{prorob}.\qed
To address problem (\ref{sec6}), firstly we  provide a lower bounding approximation for the first term in
the objective function in (\ref{sec6a}) as follows
\[
\begin{array}{lll}
\ln(1+\bR_i)&\geq& A^{(\kappa)}_i(\bR_i)
\triangleq \ds\ln(1+\Rik)+\frac{\Rik}{\Rik+1}-\frac{(\Rik)^2}{\Rik+1}\frac{1}{\bR_i}.
\end{array}
\]
Next, to obtain an upper bounding approximation for the function in the left hand side of (\ref{sec6be}) and thus to provide an inner
approximation for constraint (\ref{sec6be}), we use the following inequality
\begin{eqnarray}
\bR_{i}\pmb{p}_{j}&=&0.5(\bR_{i}+\pmb{p}_{j})^2-0.5\bR_{i}^2-0.5\pmb{p}_{j}^2\nonumber\\
&\leq&\Upsilon_{ij}^{(\kappa)}(\bR_i,\pmb{p}_j)\nonumber\\
&\triangleq&0.5(\bR_{i}+\pmb{p}_{j})^2-\Rik\bR_i+0.5(\Rik)^2
-\pjk\bp_j+0.5(\pjk)^2,\label{sec5.2R}
\end{eqnarray}
over the trust region
\begin{equation}\label{tr3}
2\bR_i\geq \Rik, 2\bp_j\geq \pjk.
\end{equation}
Then
\begin{eqnarray}
\ds\bp_i\ln\left(1+\frac{\bar{h}_{ji}\bR_{i}\bp_j}{\bar{h}_{ii}\bp_i}\right)&\leq&
\ds\bp_i\left[\ln\left(1+\frac{\bar{h}_{ji}\Rik \pjk}{\bar{h}_{ii}\pik}\right)
+\frac{1}{\frac{\bar{h}_{ii}}{\bar{h}_{ji}}+\frac{\Rik \pjk}{\pik}}
(\frac{\bR_{i}\bp_{j}}{\pmb{p}_{i}}
-\frac{\Rik\pjk}{\pik})\right]\nonumber\\
&\leq&\Phi_{ij}^{(\kappa)}\left(\bR_i,\pmb{p}_j,\pmb{p}_i\right)\nonumber\\
&\triangleq&\bp_i\ln\left(1+\frac{\bar{h}_{ji}\Rik \pjk}{\bar{h}_{ii}\pik}\right)\\
&&\ds+\frac{1}{\frac{\bar{h}_{ii}}{\bar{h}_{ji}}+\frac{\Rik \pjk}{\pik}}
\left(\Upsilon_{ij}^{(\kappa)}(\bR_i,\pmb{p}_j)
-\ds\frac{\Rik \pjk}{\pik}\bp_i\right).\label{sec5.3R}
\end{eqnarray}
Initialized from a feasible $(p^{(0)}, R^{(0)}, r^{(0)})$ for (\ref{sec6}) we solve the following convex program at
the $\kappa$-th iteration to generate the next iterative point $(p^{(\kappa+1)},R_l^{(\kappa+1)}, r_u^{(\kappa+1)})$:
\begin{subequations}\label{sec5kR}
\begin{eqnarray}
\max_{\pmb{w},\br}\min_{i=1,...,M} [A_i^{(\kappa)}(\bR_i)-a_i^{(\kappa)}(\br_i)]\quad\mbox{s.t}\quad (\ref{sec4b}), {\color{black}(\ref{sec5c})}, (\ref{tr1}),  (\ref{sec5kb}), (\ref{sec6c}),  (\ref{tr3})\label{sec5kRa}\\
\ds\bp_i\bar{h}_{ii}\left[\delta\ln(1-\epsilon_c)-1\right]+\sigma_i^2\bR_i+\ds\sum_{j\neq i}\bar{h}_{ji}\Upsilon_{ij}^{(\kappa)}(\bR_i,\bp_j)+
\delta\bar{h}_{ii}\sum_{j\neq i}^M\Phi_{ij}^{(\kappa)}(\bR_i,\pmb{p}_j,\pmb{p}_i)\leq 0,
\label{sec5kRb}\\
\ i=1,...,M.\nonumber
\end{eqnarray}
\end{subequations}
At the same $\kappa$-th iteration, $r_i^{(\kappa+1)}$ is found from solving (\ref{bi1})
by bisection on $[0,r_{u,i}^{(\kappa+1)}]$ such that (\ref{bi2}),
while $R_i^{(\kappa+1)}$ is found from solving
\begin{equation}\label{nonlin1}
\zeta_i(\mathbf{R}_i)=0, i=1,\dots, M,
\end{equation}
for the increasing function
\[
\zeta_i(\mathbf{R}_i)\triangleq \delta\ln(1-\epsilon_c)-1+\ds\frac{\bR_{i}(\sigma_i^2+\sum_{j\neq i}\bar{h}_{ji}p_j^{(\kappa+1)})}{\bar{h}_{ii}p _{i}^{(\kappa+1)}}+
\delta\sum_{j\neq i}^M\ln\left(1+\frac{\bar{h}_{ji}\bR_{i}p_{j}^{(\kappa+1)}}{\bar{h}_{ii}p_{i}^{(\kappa+1)}}\right),
\]
by bisection on $[R_{l,i}^{(\kappa+1)}, R_{u,i}]$ with  $\zeta_i(R_{u,i})>0$ such that
\begin{equation}\label{bi3}
-\epsilon_b\leq g_i(R_i^{(\kappa+1)})\leq 0.
\end{equation}
$R_{u,i}$ can be easily located: initialized by $R_i=2R_{l,i}^{(\kappa+1)}$ and check $\zeta_i(R_i)$.
If $\zeta_i(R_i)>0$ then we are done. Otherwise reset $R_i\leftarrow 2R_i$ until $\zeta_i(R_i)>0$.
Intuitively, taking $R_{u,i}=2\bar{h}_{ii}p_i^{(\kappa+1)}/(\sigma_i^2+\sum_{j\neq i}\bar{h}_{ji}p_j^{(\kappa+1)})$
will work.

In summary,  we propose in Algorithm \ref{alg3}
a path-following computational procedure for the maximin secrecy throughput optimization problem (\ref{sec6}),
which at least converges to its locally optimal solution satisfying the first order necessary optimality condition.
\begin{algorithm}
\caption{Path-following algorithm for maximin secrecy throughput optimization } \label{alg3}
\begin{algorithmic}
\STATE \textbf{Initialization}: Set $\kappa=0$. Choose an initial feasible point
$(p^{(0)}, R^{(0)}, r^{(0)})$ for (\ref{sec6}) and calculate $R_{\min}^{(0)}$ as the value of the objective function in
(\ref{sec6}) at $(p^{(0)}, R^{(0)}, r^{(0)})$.
\REPEAT \STATE $\bullet$ Set
$\kappa=\kappa+1$.
\STATE $\bullet$ Solve the
convex optimization problem \eqref{sec5kR} to obtain the
solution $(p^{(\kappa)}, R_{l}^{(\kappa)}, r_u^{(\kappa)})$.
\STATE $\bullet$ Solve the nonlinear equations (\ref{bi1}) to obtain the roots $r_i^{(\kappa)}$.
\STATE $\bullet$ Solve the nonlinear equations (\ref{nonlin1}) to obtain the roots $R_i^{(\kappa)}$.
\STATE $\bullet$ Calculate  $R_{\min}^{(\kappa)}$ as the value of the objective function in (\ref{sec6}) at
$(p^{(\kappa)}, R^{(\kappa)}, r^{(\kappa)})$.
 \UNTIL{$\frac{R_{\min}^{(\kappa)}-R_{\min}^{(\kappa-1)})}{R_{\min}^{(\kappa-1)}} \leq
 \epsilon_{\rm tol}$}.
\end{algorithmic}
\end{algorithm}

An initial  feasible $(p^{(0)},R^{(0)}, r^{(0)})$ can be easily found as follows: taking any $p^{(0)}$ feasible
to the power constraint (\ref{sec4b}) to find $R^{(0)}$ and $r^{(0)}$ from solving
\[
\begin{array}{r}
\zeta_i(\bR_i)\triangleq \delta\ln(1-\epsilon_c)-1+\ds\frac{\bR_{i}(\sigma_i^2+\sum_{j\neq i}\bar{h}_{ji}p_j^{(0)})}{\bar{h}_{ii}p_{i}^{(0)}}+\delta
\sum_{j\neq i}^M\ln\left(1+\frac{\bar{h}_{ji}\bR_{i}p_{j}^{(0)}}{\bar{h}_{ii}p_{i}^{(0)}}\right)= 0,\\
i=1,...,M,
\end{array}
\]
by bisection on $[0, 2\bar{h}_{ii}p_i^{(0)}/(\sigma_i^2+\sum_{j\neq i}\bar{h}_{ji}p_j^{(0)})]$,
and $r^{(0)}$ is found from solving
\[
\ln(1-\epsilon_c)+\ds\frac{\br_{i}\sigma_e^2}{\bar{h}_{ie}p_i^{(0)}}+
\sum_{j\neq i}^M\ln\left(1+\frac{\br_{i}\bar{h}_{je}p_j^{(0)}}{\bar{h}_{ie}p_i^{(0)}}\right)= 0,
\ i=1,...,M,
\]
by bisection on $[0,\bar{h}_{ie}p_i^{(0)}/\sigma_e^2]$.

Lastly, the network secure energy efficiency problem is now formulated by
\begin{subequations}\label{esec6}
\begin{eqnarray}
\ds\max_{\pmb{p},\bR,\br}\frac{\sum_{i=1}^{M} (\ln(1+\bR_i)-\ln(1+\br_i))}{\pi(\bp)}
\quad\mbox{s.t}\quad (\ref{sec4b}),  {\color{black}(\ref{sec5c})}, (\ref{sec5b}),
(\ref{rotage3}), (\ref{sec6c}), \label{esec6a}\\
\ln(1+\bR_i)-\ln(1+\br_i)\geq c_i, i=1,\dots,M.\label{esec6b}
\end{eqnarray}
\end{subequations}
To this end, we use inequality (\ref{esec5.2}) in Appendix II  to obtain
\begin{eqnarray}
\ds\frac{\ln(1+\bR_i)}{\pi(\bp)}&\geq&\tilde{A}^{(\kappa)}_i(\bR_i,\bp)\nonumber\\
&\triangleq&\ds\frac{2\ln(1+\Rik)}{\pi(\pk)}+\frac{\Rik}{\pi(\pk)(1+\Rik)}\left(1-\frac{\Rik}{\bR_i}\right)-
\frac{\ln(1+\Rik)}{\pi^2(\pk)}\pi(\bp)
\end{eqnarray}
Initialized by a feasible point $(R^{(0)}, r^{(0)}, p^{(0)})$,
at the $\kappa$-th iteration, the following convex programm is solved to generated $(R^{(\kappa+1)}, r^{(\kappa+1)}, p^{(\kappa_1)})$
\begin{subequations}\label{esec5kR}
\begin{eqnarray}
\max_{\pmb{w},\br}\sum_{i=1}^{M}\left[\tilde{A}^{(\kappa)}_i(\bR_i,\bp)+\tilde{a}^{(\kappa)}_i(\br_i,\bp)\right] \quad\mbox{s.t}\quad (\ref{sec4b}), {\color{black}(\ref{sec5c})}, (\ref{tr1}),  (\ref{sec5kb}), (\ref{sec6c}),  (\ref{tr3}),
(\ref{sec5kRb}),\label{esec5kRa}\\
A_i^{(\kappa)}(\bR_i)-a_i^{(\kappa)}(\br_i)\geq c_i, \ i=1,...,M.\label{esec5kRb}
\end{eqnarray}
\end{subequations}
It can be shown that the computational procedure that invokes the convex program (\ref{esec5kR}) to generate
the next iterative point, is path-following for (\ref{esec6}), which at least converges to its locally optimal solution satisfying the KKT conditions.\\
A point $(p^{(0)}, R^{(0)}, r^{(0)})$ is feasible for (\ref{esec6}) if and only if
$\min_{i=1,\dots,M}[f_i(R_i^{(0)})-\ln(1+r_i^{(0)})]/c_i\geq 1$ and thus can be easily located by adapting Algorithm
\ref{alg3}.

Similarly, a path-following procedure for the following maximin SEE optimization problem can be proposed
\begin{equation}\label{mseeR}
\ds\max_{\pmb{p},\br}\min_{i=1,\dots,M}\frac{f_i(\bR_i)-\ln(1+\br_i)}{\pi_i(\bp)}\quad\mbox{s.t}\quad
(\ref{sec4b}),  {\color{black}(\ref{sec5c})}, (\ref{sec5b}),
(\ref{rotage3}), (\ref{sec6c}), (\ref{esec6b}).
\end{equation}
\section{Simulation}
This section evaluates the performance of the proposed algorithms through extensive simulation. Considered in all simulation studies is a wireless network with $M=10$ transmitter-user communication pairs and noise variance $\sigma_i^2=\sigma_e^2=1$ mW \cite{WWN15}. All channels among each pair are i.i.d. complex normal random variable with zero mean and unit variance. The drain efficiency of power amplifier $1/\zeta$ is set
to be $40 \%$ and the circuit power of each transmitter is $P_c^i=5$ mW. Besides, we set
{\color{black}$\epsilon_c=0.1$ and $\epsilon_{EV}\in\{0.1, 0.6\}$}
and $\delta=0.1$.
The computation tolerance for terminating all proposed Algorithms is $\epsilon_{\rm tol}=10^{-4}$.
\vspace*{-0.3cm}
\subsection{Maximin secrecy throughput optimization}
\begin{figure}
\centering
\includegraphics[width=3.5in]{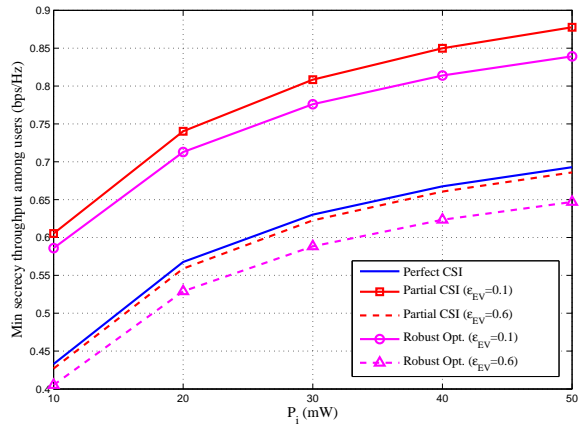}
\caption{Min secrecy throughput among users versus the transmit power budget}
\label{fig:Rate_PI}
\end{figure}
This subsection analyzes the secrecy throughput in the presence of eavesdropper. {\color{black} In what follows, we consider five cases, including ``Perfect CSI'', ``Partial CSI ($\epsilon_{EV}=0.1$)'', ``Partial CSI ($\epsilon_{EV}=0.6$)'', ``Robust Opt. ($\epsilon_{EV}=0.1$ $\epsilon_c=0.1$)'' and ``Robust Opt. ($\epsilon_{EV}=0.6$ $\epsilon_c=0.1$)''.
The terms ``Perfect CSI'', ``Partial CSI'' and ``Robust Opt.''  correspond to the scenarios discussed in Sections III, IV and V, respectively.}
Fig. \ref{fig:Rate_PI} plots the minimum secrecy throughput versus the transmit power budget $P_i$ varying from 10 to 50 mW.
As expected, it is seen that the secrecy throughput increase with the transmitted power budget $P_i$.
It is also observed that the secrecy throughput of ``Partial CSI'' with $\epsilon_{EV}=0.1$ is always better than the secrecy throughputs of others.
For $\epsilon_{EV}=0.1$, ``Partial CSI'' and ``Robust Opt.'' clearly and significantly outperform ``Perfect CSI''. However, the secrecy throughput of ``Perfect CSI'' is superior to the secrecy throughputs of ``Partial CSI'' and ``Robust Opt.'' with $\epsilon_{EV}=0.6$. This is not a surprise because according to the wiretapped throughput defined by (\ref{dwire1}) and
the throughput outage defined by (\ref{wire1})-(\ref{esec5b}), the former is seen higher than the later for small $\epsilon_{EV}$.

Table \ref{tab:rate_P_Ite} provides the average number of iterations required to solve maximin secrecy throughput optimization for the above three cases. As can be observed, our proposed algorithm converges in less than 14 iterations, on average, for all considered cases.

\begin{table}
   \centering
   \caption{Average number of iterations for maximin secrecy throughput optimization.}
   \begin{tabular}{ | c | c | c | c | c | c | }
    \hline
   $P_i$ (mW) & 10  &  20 &  30 &  40 &  50  \\
   \hline
   Perfect CSI & 8.12 & 7.63 &  7.61 &  7.71 &  8.56 \\
    \hline
   Partial CSI ($\epsilon_{EV}=0.1$) & 11.25 & 10.87 & 10.73 & 10.40 & 10.31 \\
  \hline
   Partial CSI ($\epsilon_{EV}=0.6$) & 13.12 & 12.18 & 14.92 & 12.60 & 11.68 \\
  \hline
  Robust Opt. ($\epsilon_{EV}=0.1$) & 4.20 & 4.33 & 4.20 & 3.52 & 3.35 \\
  \hline
  Robust Opt. ($\epsilon_{EV}=0.6$) & 5.18 & 4.96 & 4.82 & 4.14 & 4.90 \\
  \hline
\end{tabular}
\label{tab:rate_P_Ite}
\end{table}

\begin{figure}
\centering
\includegraphics[width=3.5in]{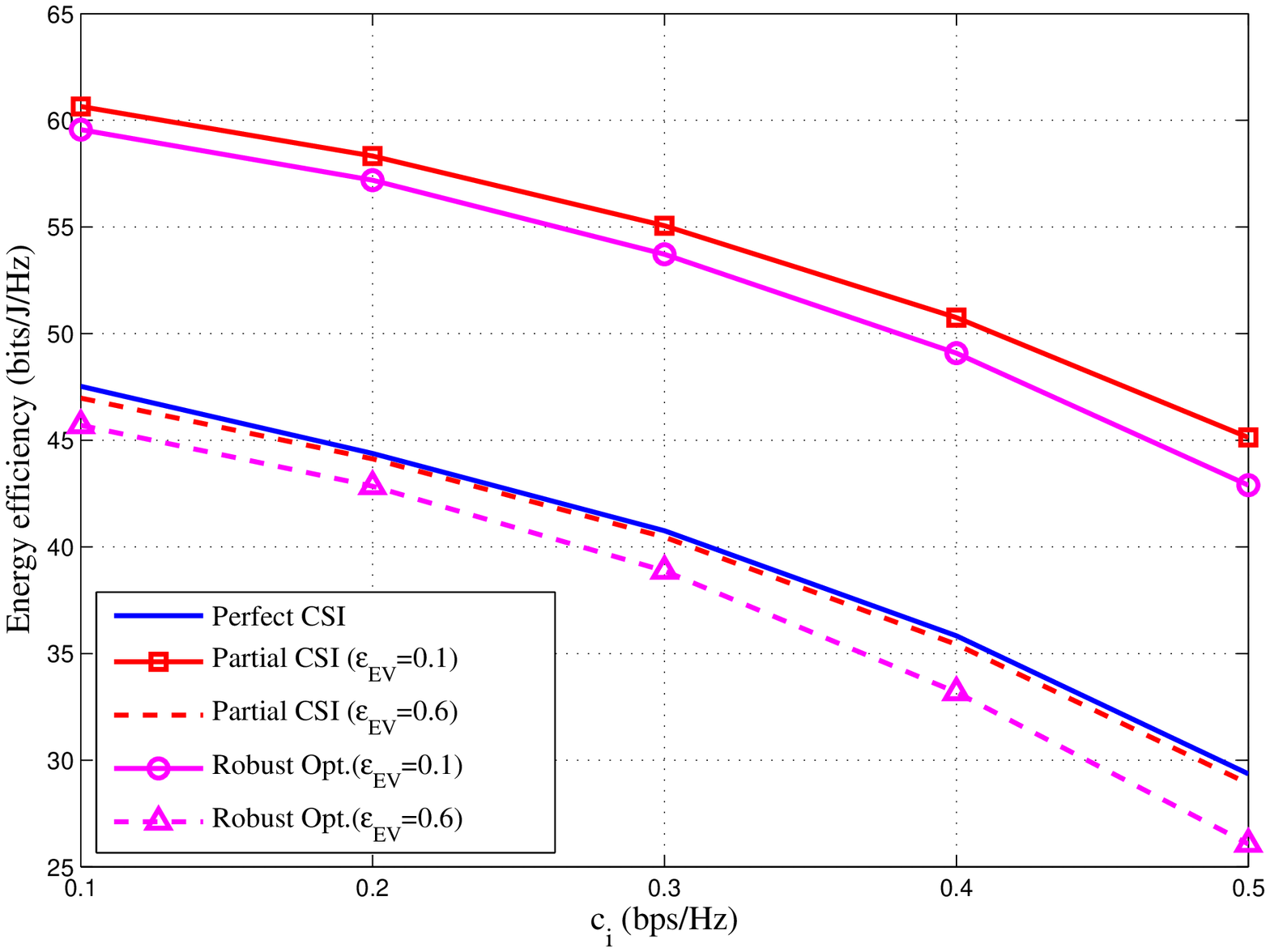}
\caption{Energy efficiency versus QoS constraint}
\label{fig:EE_c}
\end{figure}

\begin{figure*}[t]
    \centering
    \begin{minipage}[h]{0.48\textwidth}
    \centering
    \includegraphics[width=1.05 \textwidth]{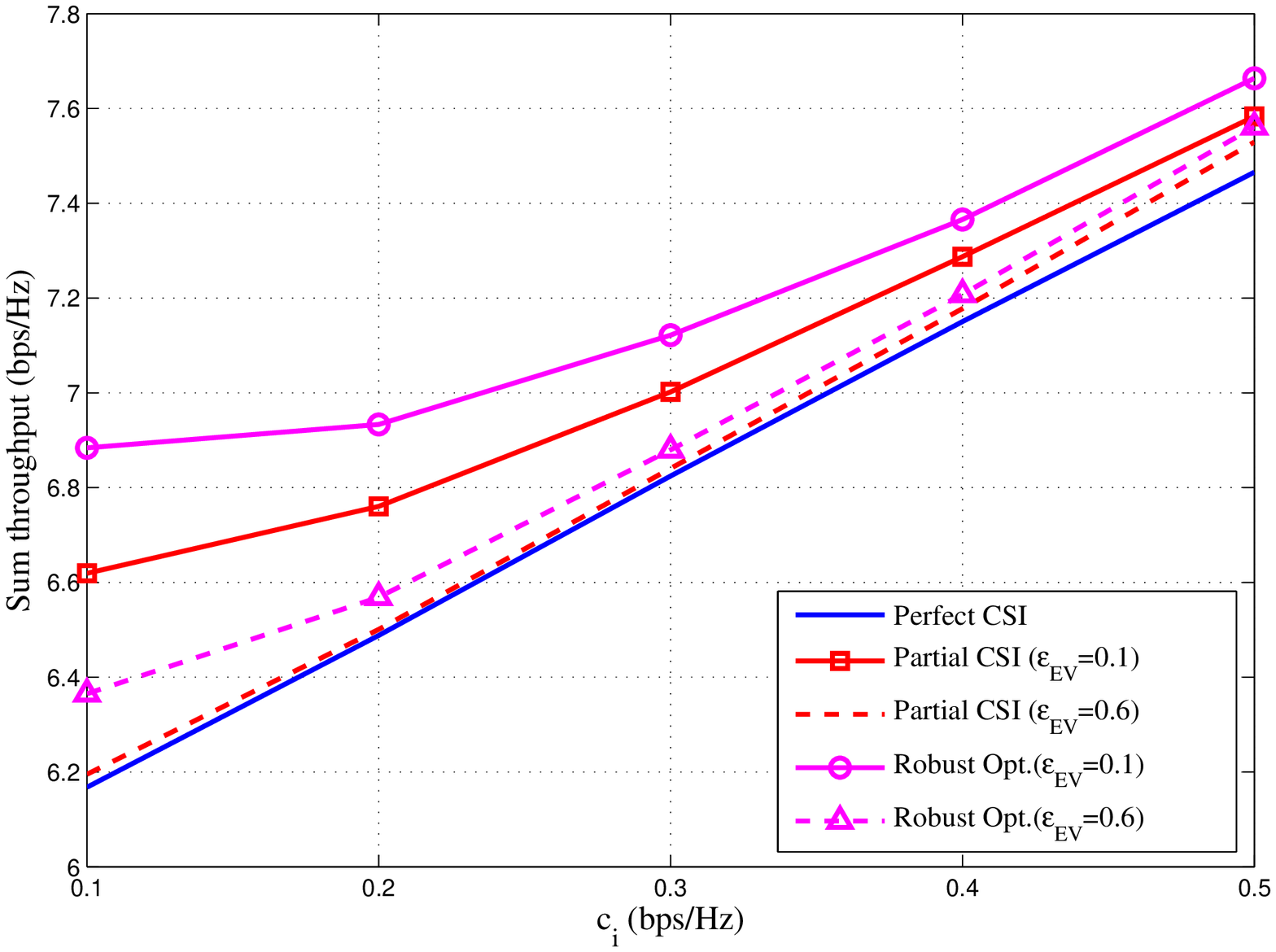}
  \caption{Sum throughput versus QoS constraint}
  \label{fig:Rate_c}
  \end{minipage}
    \hspace{0.1cm}
    \begin{minipage}[h]{0.48\textwidth}
    \centering
    \includegraphics[width=1.05 \textwidth]{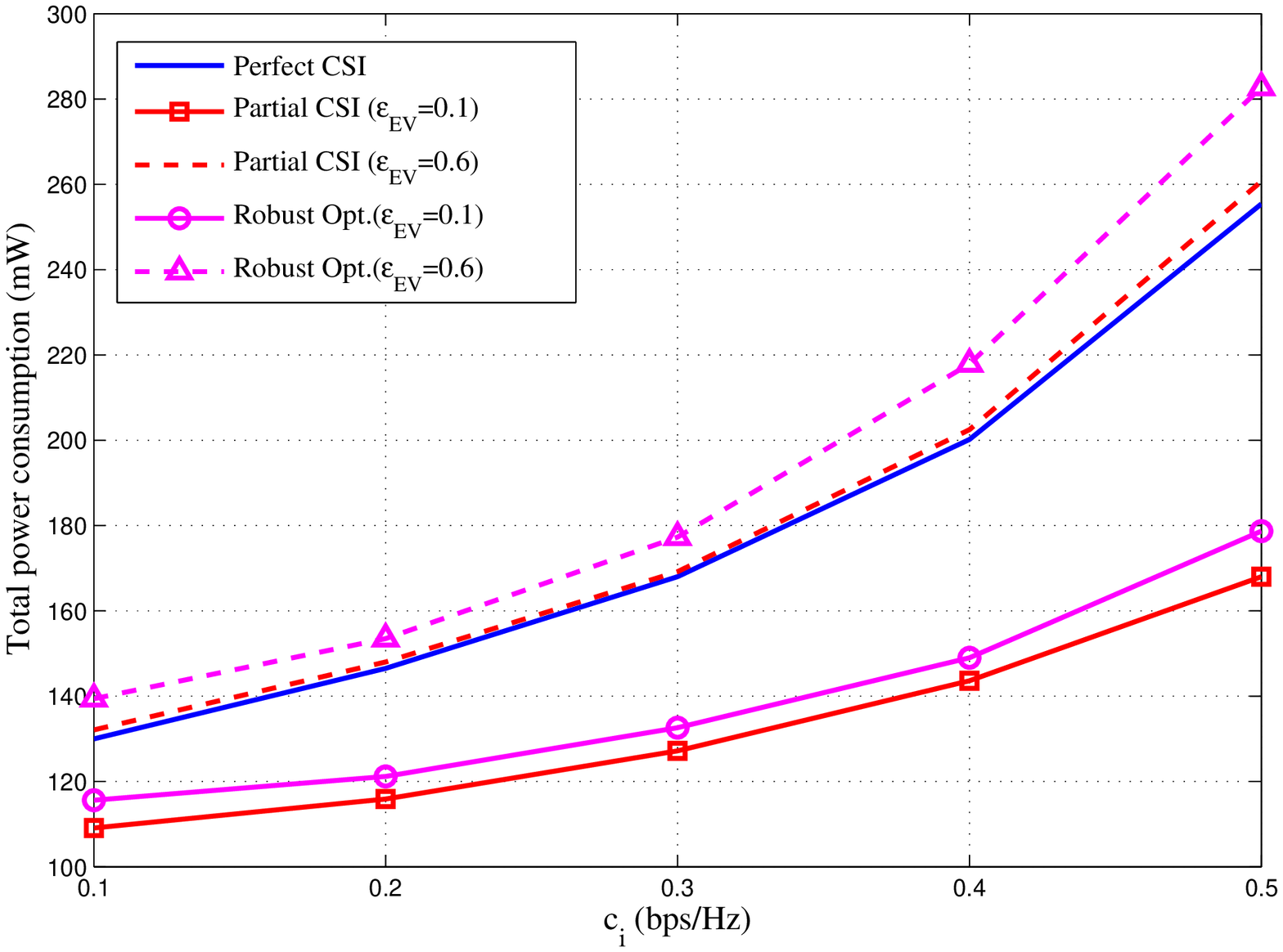}
  \caption{Total power consumption versus QoS constraint}
  \label{fig:Power_c}
  \end{minipage}
\end{figure*}

%

\vspace*{-0.3cm}
\subsection{Energy efficiency maximization}
In this subsection, we first examine the performance of EE maximization algorithm versus the QoS constraint. The transmitted power $P_i$ is
fixed at 20 mW and QoS constraint $c_i$ varies from 0.1 to 0.5 bps/Hz. It can be observed from Fig. \ref{fig:EE_c}
that the EE performance degrades as the QoS constraint $c_i$ increases. Moreover, ``Partial CSI'' with $\epsilon_{EV}=0.1$ outperforms others in terms of EE performance. To find out the impact on the sum throughput and total power consumption in EE maximization algorithm, the achieved sum throughput and the total power consumed are illustrated in Fig. \ref{fig:Rate_c} and \ref{fig:Power_c}, respectively.
It can be seen that the total power consumption increases faster than the sum throughput, which explains why
EE degrades as $c_i$ increases in Fig. \ref{fig:EE_c}.
Although the sum throughput of ``Robust Opt.'' is slightly better than ``Partial CSI'', ``Partial CSI'' consumes less power
than ``Robust Opt.''.
Table \ref{tab:EE_c_Ite} shows that our proposed EE maximization algorithm converges in less than 35 iterations, on average, in all considered cases.

\begin{table}
   \centering
   \caption{Average number of iterations for energy efficiency maximization.}
   \begin{tabular}{ | c | c | c | c | c | c | }
    \hline
   $c_i$ (bps/Hz) & 0.1  &  0.2 &  0.3 &  0.4 &  0.5  \\
   \hline
   Perfect CSI & 32.21 &  29.62 &  24.26 &  21.23  & 15.33 \\
    \hline
   Partial CSI ($\epsilon_{EV}=0.1$) & 33.73 & 33.12 & 28.75 & 25.74 & 23.25 \\
  \hline
   Partial CSI ($\epsilon_{EV}=0.6$) & 35.82 & 30.56 & 34.22 & 22.26 & 18.34  \\
  \hline
  Robust Opt. ($\epsilon_{EV}=0.1$) & 24.25  & 27.41 & 25.53  &  30.06 & 31.75  \\
  \hline
  Robust Opt. ($\epsilon_{EV}=0.6$) & 29.02  & 23.76 & 26.80 & 29.32 & 23.46  \\
  \hline
\end{tabular}
\label{tab:EE_c_Ite}
\end{table}

\begin{figure}
\centering
\includegraphics[width=3.5in]{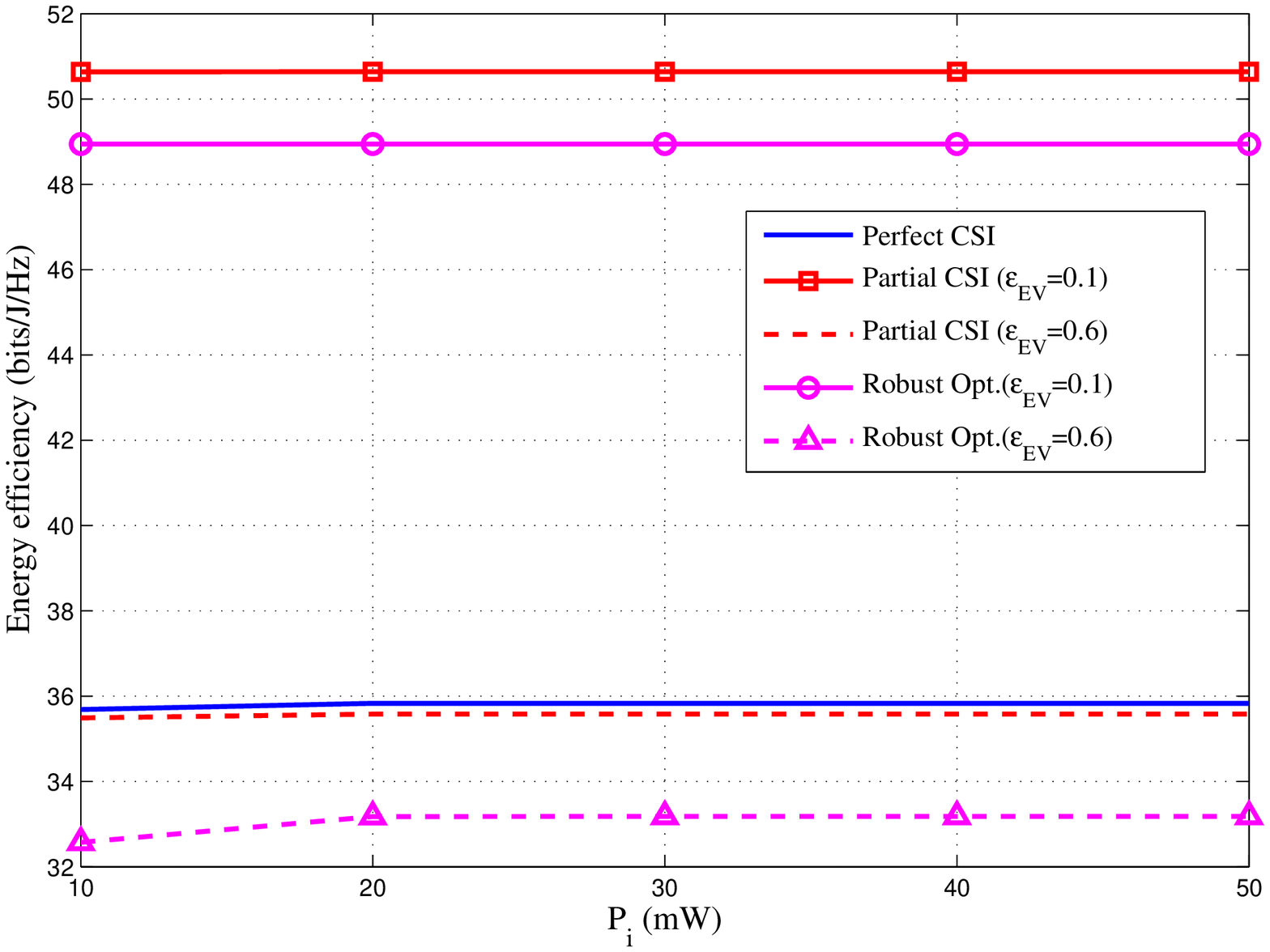}
\caption{Energy efficiency versus the transmit power budget}
\label{fig:EE_P}
\end{figure}

\begin{figure*}[t]
    \centering
    \begin{minipage}[h]{0.48\textwidth}
    \centering
    \includegraphics[width=1.05 \textwidth]{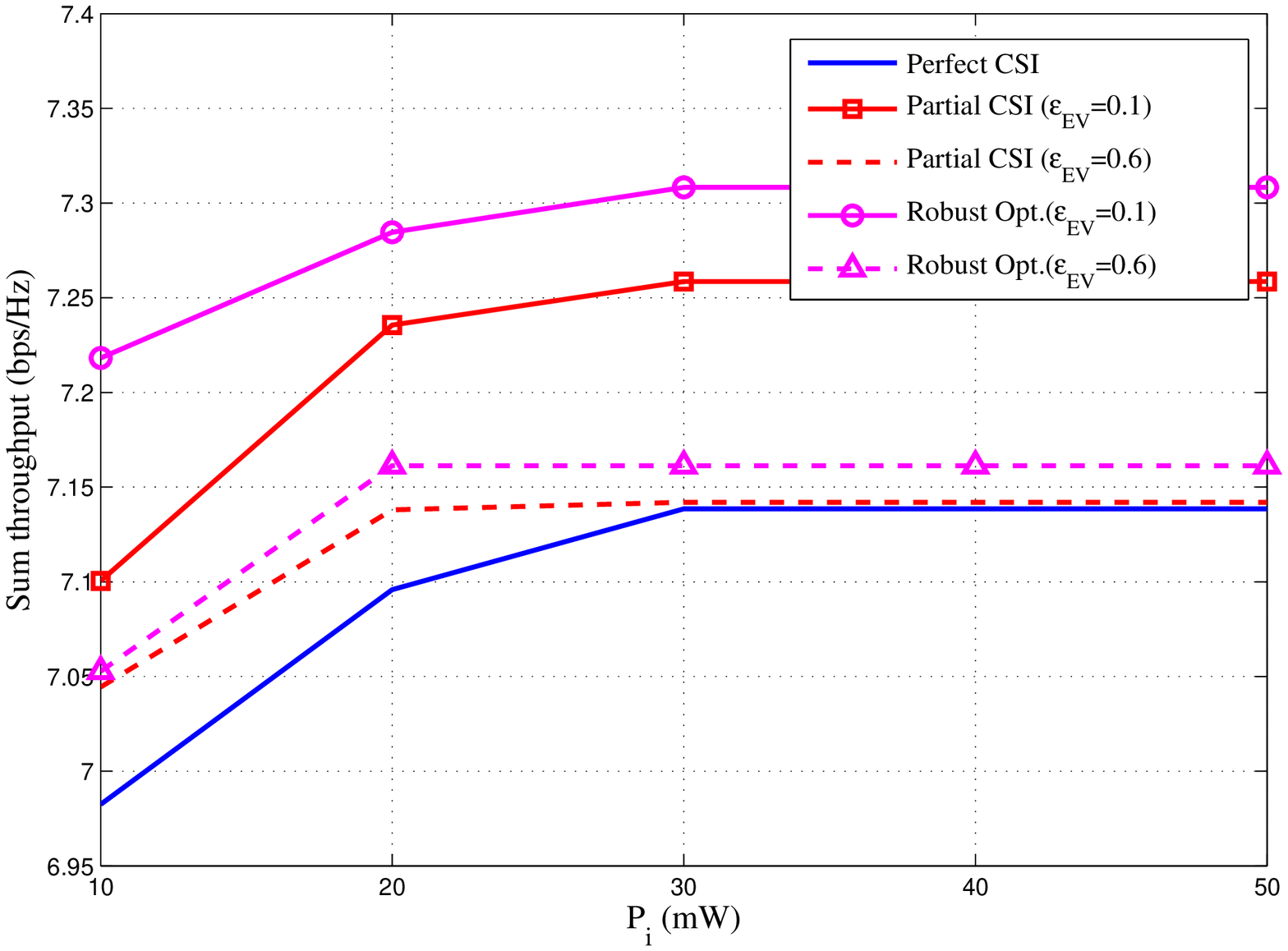}
  \caption{Sum throughput versus the transmit power budget}
  \label{fig:Rate_P}
  \end{minipage}
    \hspace{0.1cm}
    \begin{minipage}[h]{0.48\textwidth}
    \centering
    \includegraphics[width=1.05 \textwidth]{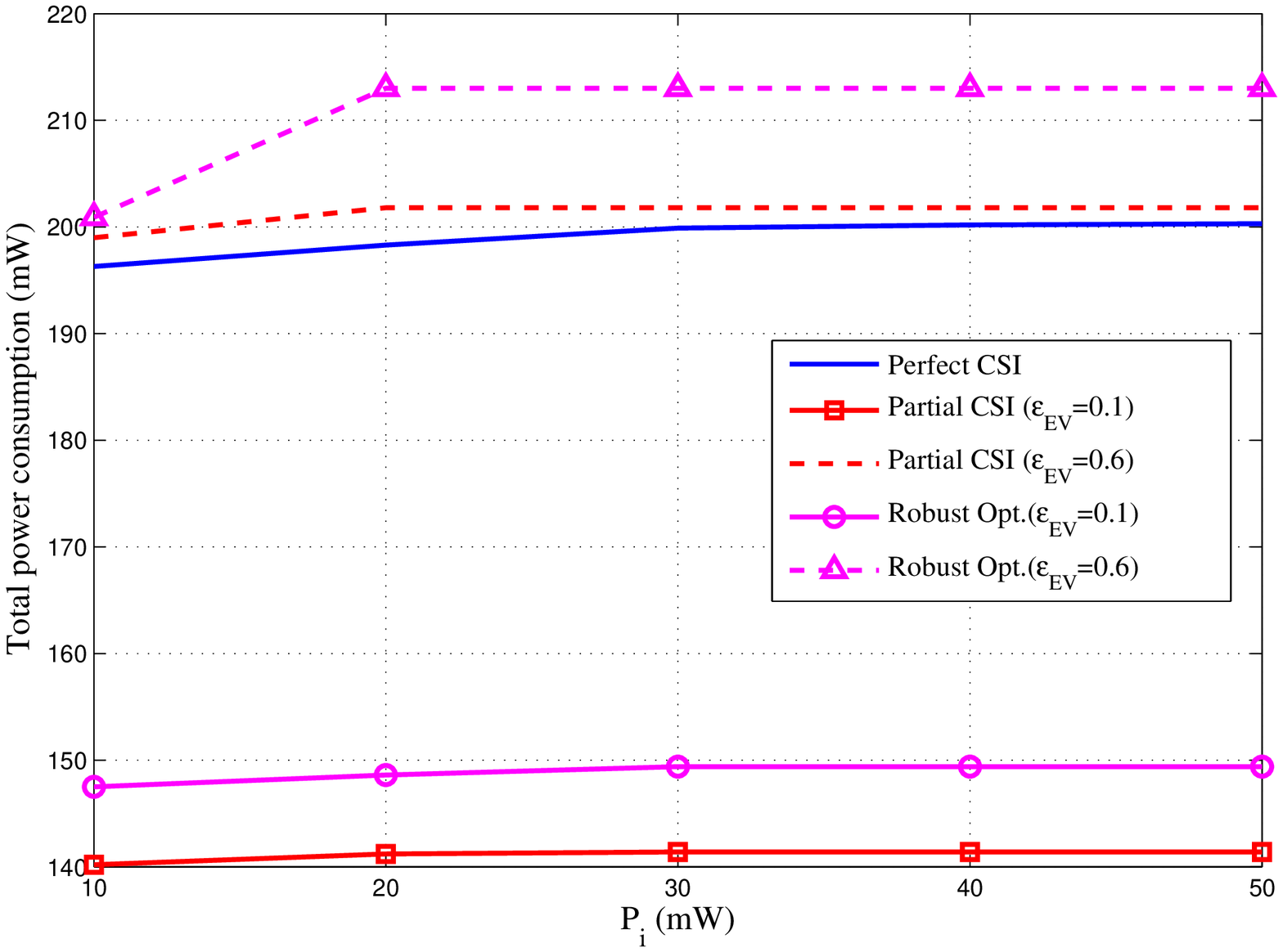}
  \caption{Total power consumption versus the transmit power budget}
  \label{fig:Power_P}
  \end{minipage}
\end{figure*}

%
%

Next, we further investigate the performance of EE versus the transmit power budget. The QoS constraint $c_i$
is fixed at 0.4 bps/Hz and $P_i$ varies from 10 to 50 mW. {\color{black} As shown in Fig. \ref{fig:EE_P}, we observe that the EE performance of ``Partial CSI'' with $\epsilon_{EV}=0.1$ clearly and significantly outperforms the optimized EE of other cases.}
Furthermore, it can be seen that the EE performances saturate when the transmit power budget
exceeds the threshold.
That is because for small transmit power budget, the denominator of EE is dominated by the circuit power and
the EE is maximized by maximization of the sum throughput in the numerator.
However, for larger transmit power budget, the denominator of EE is dominated by the actual transmit power.
When the total power consumption saturates in Fig. \ref{fig:Power_P}, both the
EE and the sum throughput accordingly saturate in Figs. \ref{fig:EE_P} and \ref{fig:Rate_P}.

\begin{figure}
\centering
\includegraphics[width=3.5in]{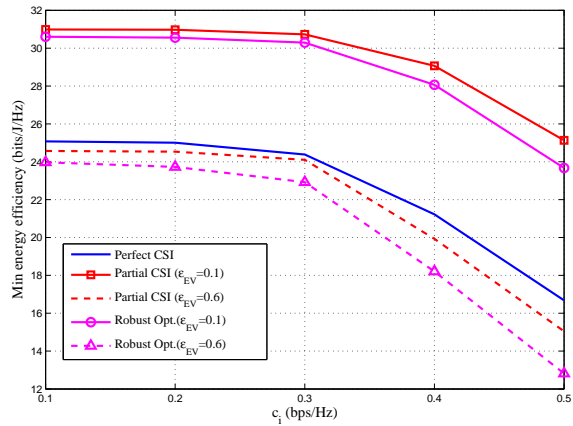}
\caption{Minimum energy efficiency versus the QoS constraint}
\label{fig:Maxmin_EE_c}
\end{figure}

\begin{figure*}[t]
    \centering
    \begin{minipage}[h]{0.48\textwidth}
    \centering
    \includegraphics[width=1.05 \textwidth]{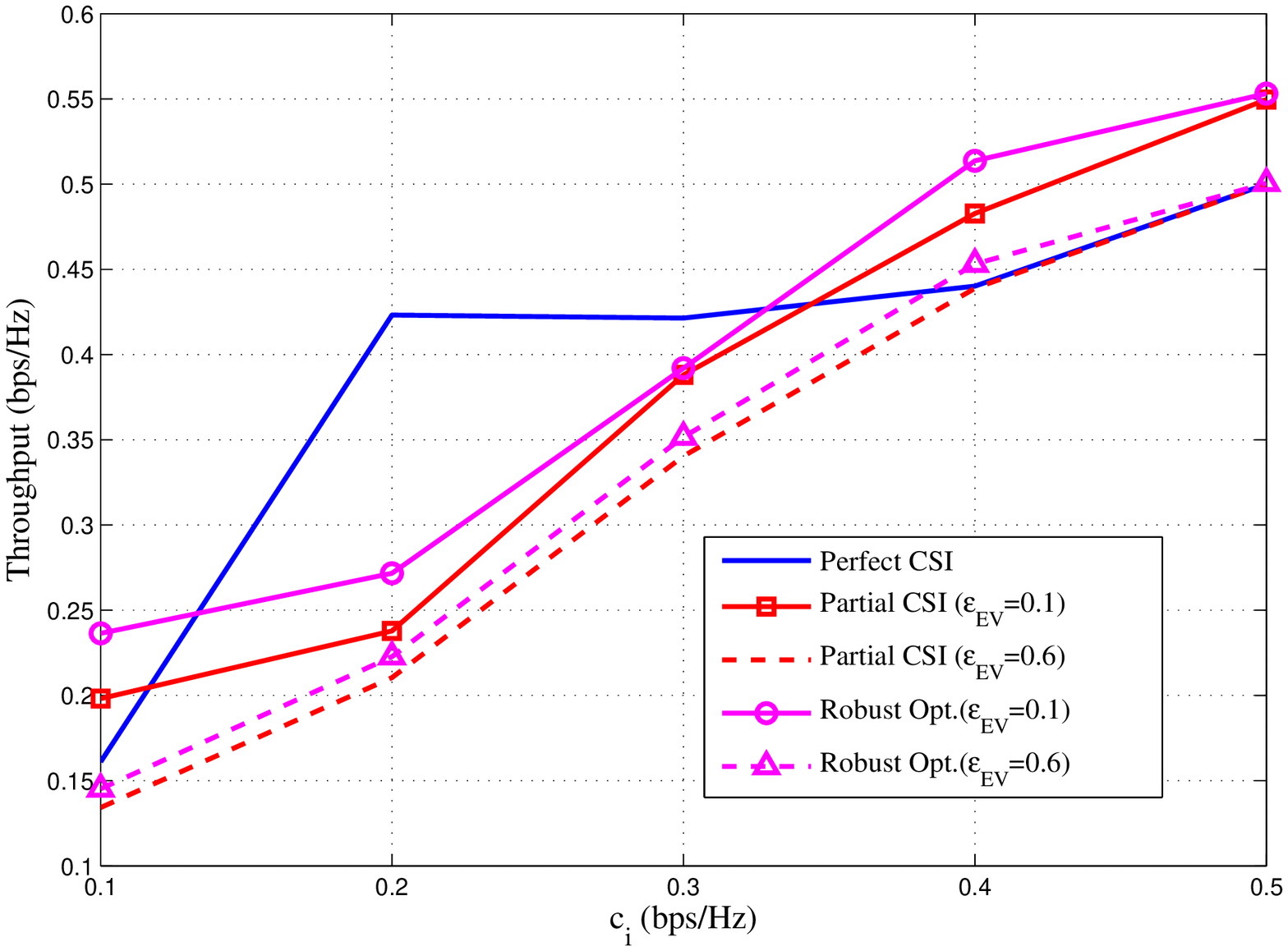}
  \caption{Throughput versus the QoS constraint}
  \label{fig:Maxmin_Rate_c}
  \end{minipage}
    \hspace{0.1cm}
    \begin{minipage}[h]{0.48\textwidth}
    \centering
    \includegraphics[width=1.05 \textwidth]{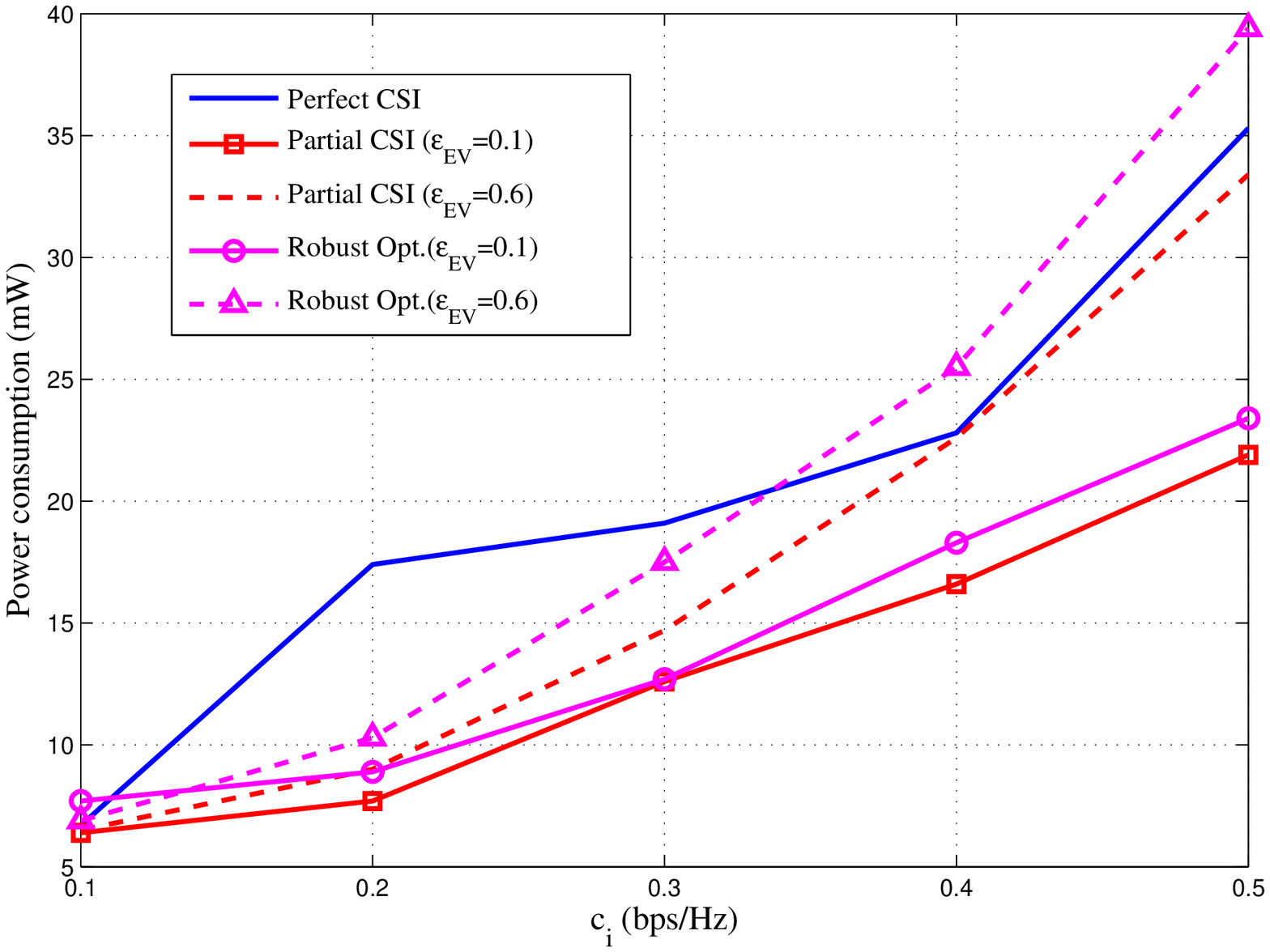}
  \caption{Power consumption versus the QoS constraint}
  \label{fig:Maxmin_Power_c}
  \end{minipage}
\end{figure*}

%

\vspace*{-0.3cm}
\subsection{Maxmin energy efficiency optimization}
In this subsection, we aim to maximize the minimum EE performance.
Firstly, Fig. \ref{fig:Maxmin_EE_c} plots the maximized minimum EE versus QoS constraint. The transmitted power $P_i$ is
fixed at 20 mW and QoS constraint $c_i$ varies from 0.1 to 0.5 bps/Hz.
{\color{black}It can be seen that the optimized minimum EE decreases with increasing $c_i$ and the EE performance of ``Partial CSI'' with $\epsilon_{EV}=0.1$ is always better than the optimized EE of other cases.
Furthermore, it is also observed that for $\epsilon_{EV}=0.1$ ``Partial CSI'' and ``Robust Opt.'' outperform ``Perfect CSI'' in terms of EE performance, while ``Perfect CSI'' is superior to ``Partial CSI'' and ``Robust Opt.'' for $\epsilon_{EV}=0.6$.}
The corresponding throughput and power consumption
are plotted in Fig. \ref{fig:Maxmin_Rate_c} and \ref{fig:Maxmin_Power_c}, respectively.
Table \ref{tab:Maxmin_EE_c_Ite} shows that maximin EE optimization converges in less than 33 iterations, on average, in all considered cases.

\begin{table}
   \centering
   \caption{Average number of iterations for maximin energy efficiency optimization.}
   \begin{tabular}{ | c | c | c | c | c | c | }
    \hline
   $c_i$ (bps/Hz) & 0.1  &  0.2 &  0.3 &  0.4 &  0.5  \\
   \hline
   Perfect CSI & 32.42 &  30.35 & 31.61 &  29.23 & 22.25 \\
    \hline
   Partial CSI ($\epsilon_{EV}=0.1$) & 21.86 & 22.13 & 20.42 & 20.82 & 30.23 \\
  \hline
   Partial CSI ($\epsilon_{EV}=0.6$) & 23.66 & 25.02 & 22.68 & 33.30 & 29.34 \\
  \hline
  Robust Opt. ($\epsilon_{EV}=0.1$) & 16.05 &  23.27 & 23.36 & 31.15 & 18.62  \\
  \hline
  Robust Opt. ($\epsilon_{EV}=0.6$) & 20.78  &  26.12  & 31.24  & 27.46  &  23.92  \\
  \hline
\end{tabular}
\label{tab:Maxmin_EE_c_Ite}
\end{table}

\begin{figure}
\centering
\includegraphics[width=3.5in]{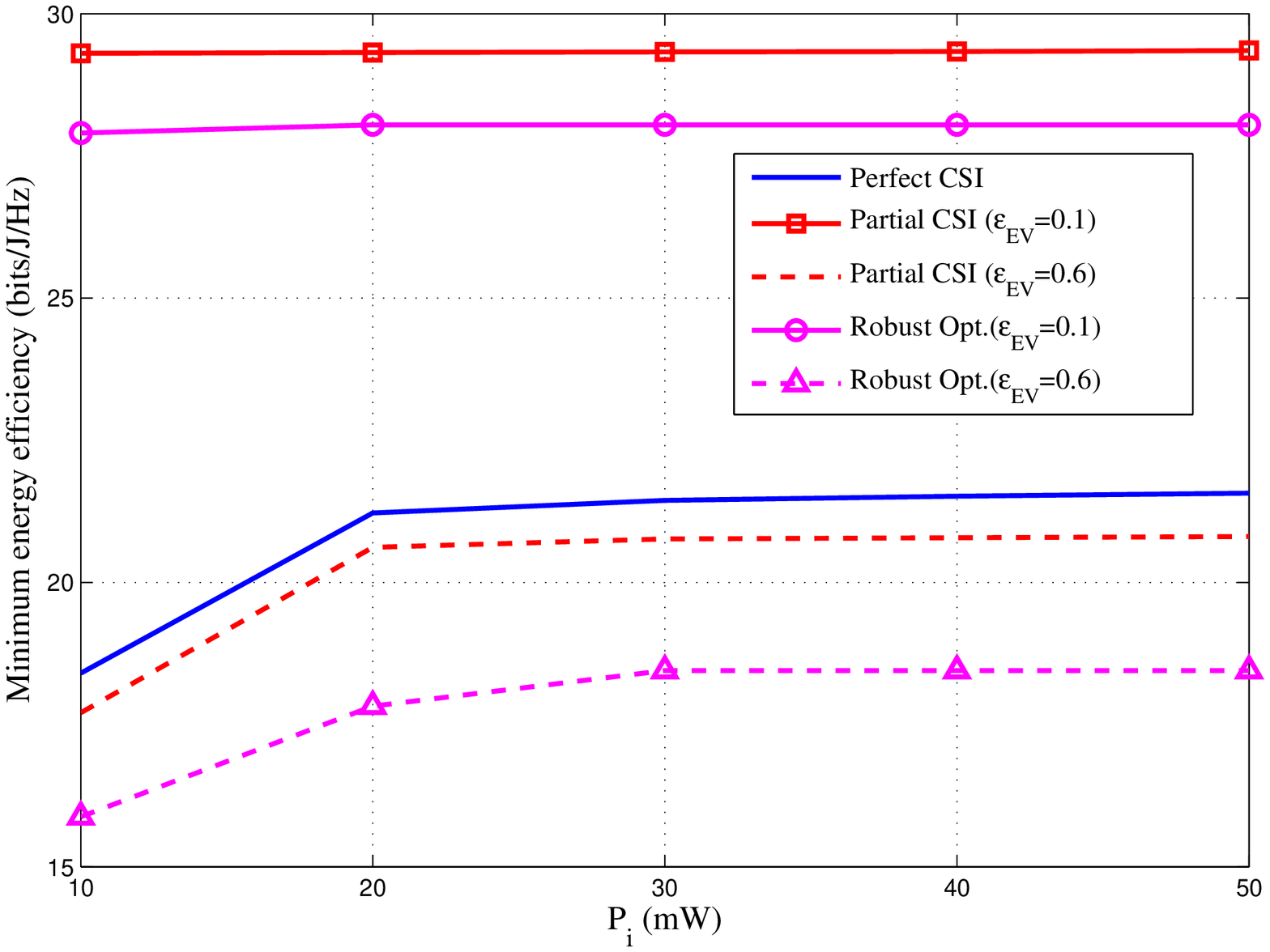}
\caption{Minimum energy efficiency versus the transmit power budget}
\label{fig:Maxmin_EE_P}
\end{figure}

\begin{figure*}[t]
    \centering
    \begin{minipage}[h]{0.48\textwidth}
    \centering
    \includegraphics[width=1.05 \textwidth]{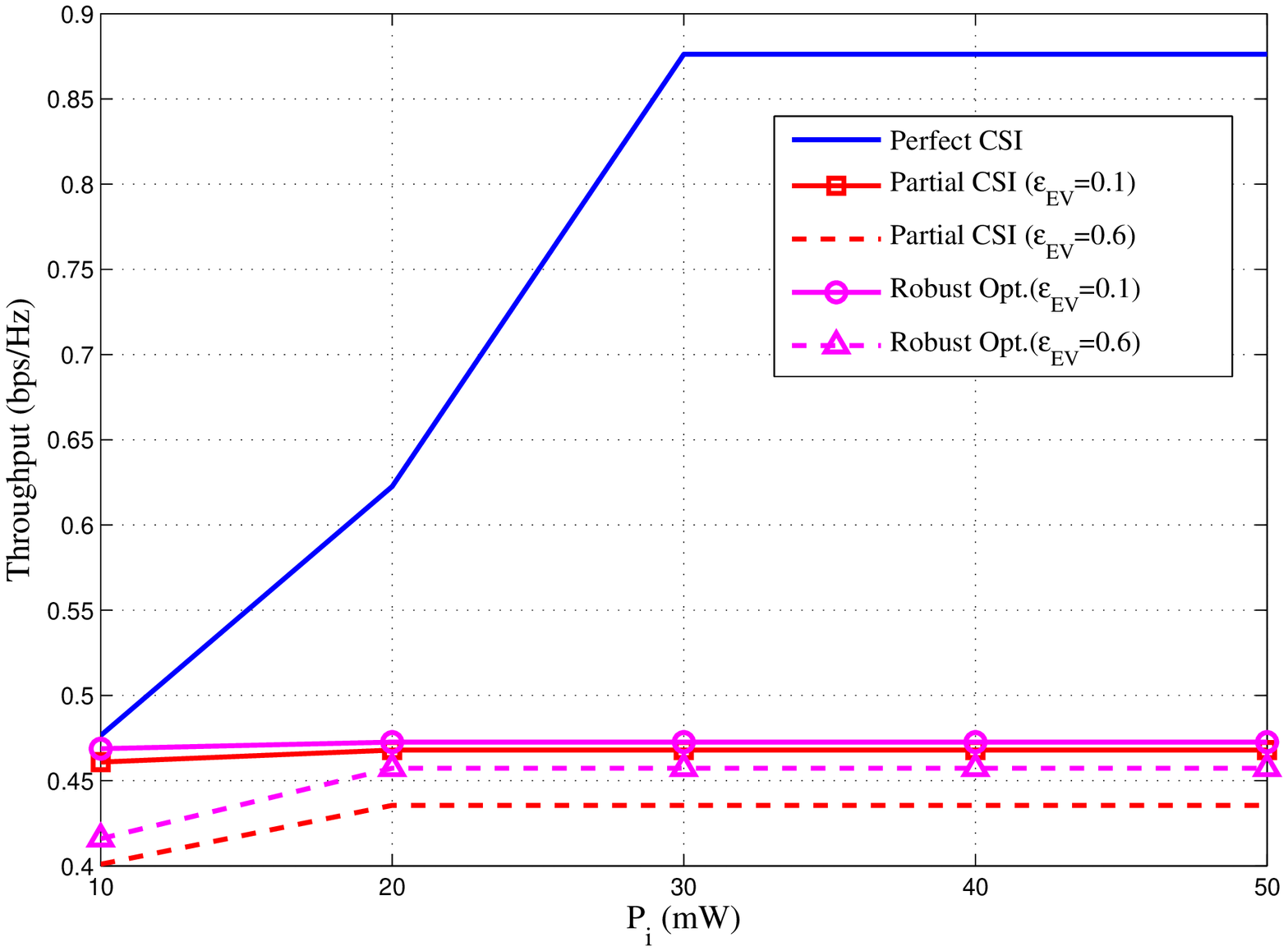}
  \caption{Throughput versus the transmit power budget}
  \label{fig:Maxmin_Rate_P}
  \end{minipage}
    \hspace{0.1cm}
    \begin{minipage}[h]{0.48\textwidth}
    \centering
    \includegraphics[width=1.05 \textwidth]{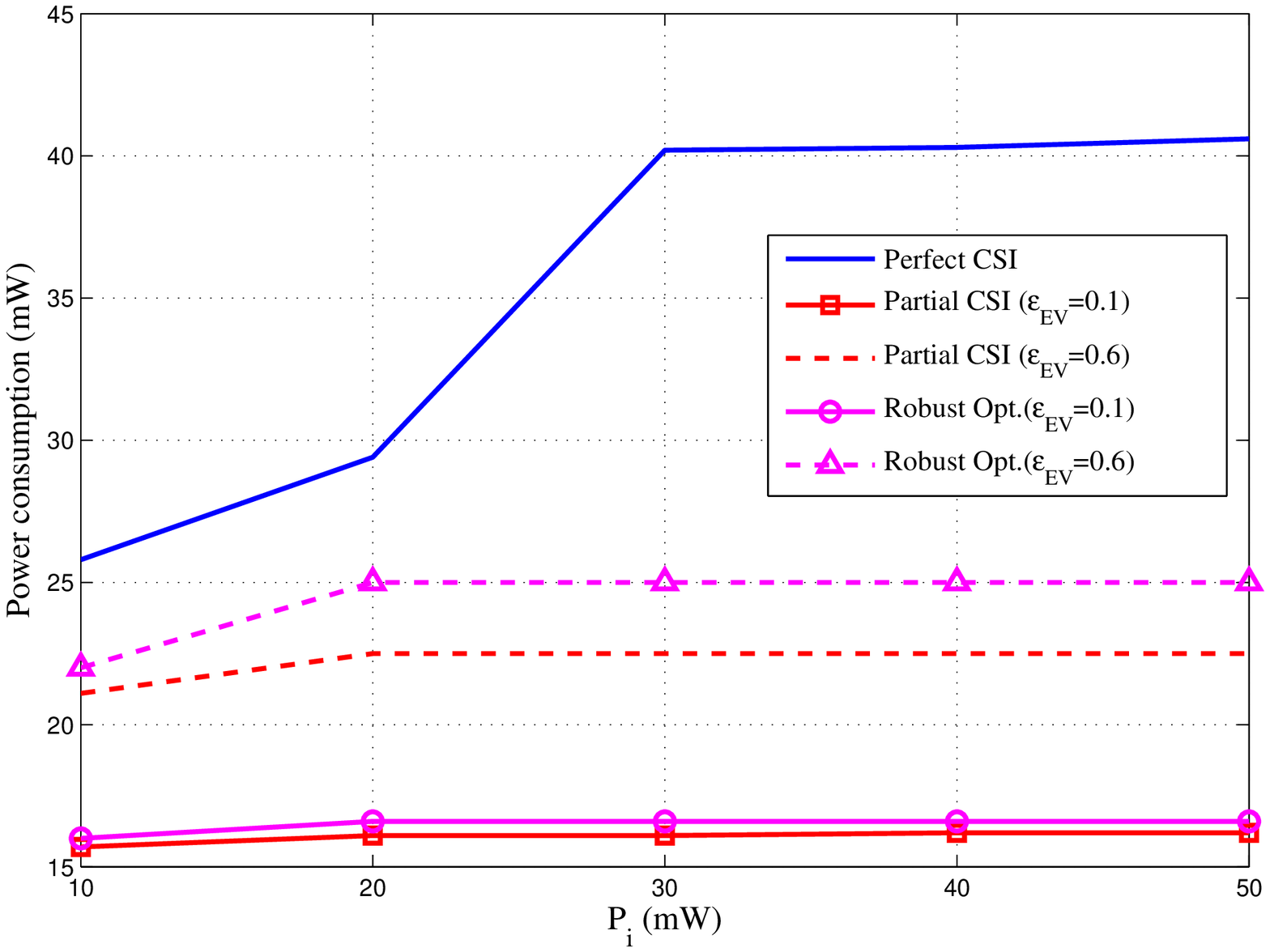}
  \caption{Power consumption versus the transmit power budget}
  \label{fig:Maxmin_Power_P}
  \end{minipage}
\end{figure*}

%

Next, we investigate the maximin EE performance versus the transmit power budget. The QoS constraint $c_i$
is fixed at 0.4 bps/Hz and $P_i$ varies from 10 to 50 mW.
The minimum EE performance versus the transmit power budget is illustrated in Fig. \ref{fig:Maxmin_EE_P}. Again, we observe that
the optimized minimum EE saturates when the transmit power is larger than some threshold. This is due to the fact that
under small transmit power regime, the EE is maximized by maximizing the throughput in the numerator.
When the transmit power is large enough to obtain the optimized EE, both throughput and power consumption accordingly saturate in Figs. \ref{fig:Maxmin_Rate_P} and \ref{fig:Maxmin_Power_P}.
\section{Conclusions}
We have considered the problem of power allocation to maximize the worst links's secrecy throughput or the network's
secure energy efficiency under various scenarios of available channel state information.
We have further proposed path-following algorithms tailored for each of the considered scenarios. Finally, we have provided
 simulations to show the efficiency of the proposed algorithms. Extensions to beamforming in multi-input single-output
 (MISO) interference networks are under current investigation.
\section*{Appendix I: outage probability fundamental}
Recall a probability distribution $\chi$ is called an {\it exponential distribution}
if its probability density function (pdf) is in form   $\lambda e^{-\lambda x}$ with
$\lambda >0$. It is immediate to check that $\Prob(\chi\geq t)=e^{-\lambda t}$ and
$\bbE[\chi]=1/\lambda$.
Recall the following result \cite[(15)]{KB02}.
\begin{myth}\label{thop1} Suppose $z_1,\cdots, z_n$ are independent exponentially distributed random variables
with $\bbE(z_i)=1/\lambda_i$. Then for deterministic $p_i>0$, $i=1,\cdots, n$:
\begin{equation}\label{ap1}
\Prob(p_1z_1\leq \sum_{i=2}^np_iz_i)=1-\prod_{i=2}^n\frac{1}{1+(\lambda_1/p_1)/(\lambda_i/p_i)}.
\end{equation}
\end{myth}
It follows from (\ref{ap1}) that
\begin{equation}\label{ap2}
\Prob(p_1z_1> c+\sum_{i=2}^np_iz_i)=e^{-\lambda_1 c/p_1}\prod_{i=2}^n\frac{1}{1+(\lambda_1/p_1)/(\lambda_i/p_i)}
\end{equation}
and
\begin{eqnarray}
\ds\Prob(\frac{p_1z_1}{\ds\sum_{i=2}^np_iz_i+\sigma}> \gamma)&=&
\ds\Prob(p_1z_1> \ds\sum_{i=2}^n\gamma p_iz_i+\gamma\sigma)\\
&=&\ds e^{-\lambda_1\gamma\sigma/p_1}\prod_{i=2}^n\frac{1}{1+\gamma(\lambda_1/p_1)/(\lambda_i/p_i)}\label{ap2.1}
\end{eqnarray}
Sometimes, it is also more convenient to write (\ref{ap1}), (\ref{ap2}) and (\ref{ap2.1}) in terms of
means $\bar{\lambda}_i=1/\lambda_i$ of $z_i$
as
\begin{eqnarray}
\Prob(p_1z_1\leq \sum_{i=2}^np_iz_i)=1-\prod_{i=2}^n\frac{p_1\bar{\lambda}_1}{p_1\bar{\lambda}_1
+p_i\lambda_i},\label{ap1e}\\
\Prob(p_1z_1> c+\sum_{i=2}^np_iz_i)=e^{-c/p_1\bar{\lambda}_1}\prod_{i=2}^n\frac{p_1\bar{\lambda}_1}{p_1\bar{\lambda}_1
+p_i\bar{\lambda}_i},\label{ap2e}\\
\ds\Prob(\frac{p_1z_1}{\sum_{i=2}^np_iz_i+\sigma}> \gamma)=
\ds e^{-\gamma\sigma/p_1\bar{\lambda}_1}\prod_{i=2}^n\frac{p_1\bar{\lambda}_1}{p_1\bar{\lambda}_1
+\gamma p_i\bar{\lambda}_i}
\end{eqnarray}
\begin{myth}\label{thop2}
For given $\varepsilon>0$, define
\begin{equation}\label{ap5}
r_{\max}\triangleq \max \{ r\ :\ \Prob(\frac{p_1z_1}{\sum_{i=2}^np_iz_i+\sigma^2})<r)\leq \varepsilon\}
\end{equation}
Then $r_{\max}$ is the unique positive root of the nonlinear equation
\begin{equation}\label{root1}
\ln(1-\varepsilon)+ \ds\frac{r\sigma^2}{p_1\bar{\lambda}_1}+\sum_{i=2}^n\ln(1+
\frac{rp_i\bar{\lambda}_i}{p_1\bar{\lambda}_1})=0.
\end{equation}
\end{myth}
\Prf Applying (\ref{ap2e}) yields
\begin{eqnarray}
\ds\Prob(\frac{p_1z_1}{\sum_{i=2}^np_iz_i+\sigma^2}<r)&=&\ds\Prob(p_1z_1<r(\ds\sum_{i=2}^np_iz_i+\sigma^2))\nonumber\\
&=&1-e^{-r\sigma^2/p_1\bar{\lambda}_1}\ds\prod_{i=2}^n\frac{p_1\bar{\lambda}_1}{p_1\bar{\lambda}_1
+rp_i\bar{\lambda}_i}.\label{ap6}
\end{eqnarray}
Therefore,
\begin{eqnarray}
\Prob(\frac{p_1z_1}{\sum_{i=2}^np_iz_i+\sigma^2})<r)\leq \varepsilon&\Leftrightarrow&1-e^{-r\sigma^2/p_1\bar{\lambda}_1}\ds\prod_{i=2}^n\frac{p_1\bar{\lambda}_1}{p_1\bar{\lambda}_1
+rp_i\bar{\lambda}_i}\leq \varepsilon\nonumber\\
&\Leftrightarrow&\ln(1-\varepsilon)+ \ds\frac{r\sigma^2}{p_1\bar{\lambda}_1}+\sum_{i=2}^n\ln(1+
\frac{rp_i\bar{\lambda}_i}{p_1\bar{\lambda}_1})\leq 0.\label{ap7}
\end{eqnarray}
By noticing that the function in the left hand side (LHS) of (\ref{ap7}) is increasing in $r$, we arrive at (\ref{root1}).\qed
\begin{myth}\label{thop3}
Suppose $\bar{z}_i>0$, $p_i>0$, $\delta>0$ and $\sigma>0$
are deterministic values, while $\tilde{z}_i$ are  independent exponential distributions. For $\varepsilon>0$, define
\begin{equation}\label{ap8}
r_{p}\triangleq \max\ \{r\ :\
\Prob(\frac{p_1\bar{z}_1(1+\delta \tilde{z}_1)}{\sum_{i=2}^np_i\bar{z}_i(1+\delta \tilde{z}_i)+\sigma^2}<r)\leq \varepsilon
\}.
\end{equation}
Then $r_p$ is the unique positive root of the nonlinear equation
\begin{equation}\label{root2}
\delta \ln(1-\varepsilon)+\frac{r(\sigma^2+\sum_{i=2}^np_i\bar{z}_i)-p_1\bar{z}_1}{\bar{z}_1p_1}+\delta
\sum_{i=2}^n\ln(1+\frac{r\bar{z}_ip_i}{\bar{z}_1p_1})=0.
\end{equation}
\end{myth}
\Prf Using (\ref{ap7}) yields
\begin{eqnarray}
\Prob(\frac{p_1\bar{z}_1(1+\delta \tilde{z}_1)}{\sum_{i=2}^np_i\bar{z}_i(1+\delta \tilde{z}_i)+\sigma^2}<r)\leq \varepsilon
&\Leftrightarrow&\nonumber\\
\ln(1-\varepsilon)+\frac{r(\sigma^2+\sum_{i=2}^np_i\bar{z}_i)-p_1\bar{z}_1}{\bar{z}_1p_1\delta}+
\sum_{i=2}^n\ln(1+\frac{r\bar{z}_ip_i}{\bar{z}_1p_1})\leq 0&\Leftrightarrow\label{ap9}\\
\delta \ln(1-\varepsilon)+\frac{r(\sigma^2+\sum_{i=2}^np_i\bar{z}_i)-p_1\bar{z}_1}{\bar{z}_1p_1}+\delta
\sum_{i=2}^n\ln(1+\frac{r\bar{z}_ip_i}{\bar{z}_1p_1})\leq
0.\label{ap10}&&
\end{eqnarray}
Again, by noticing that the function in the LHS of (\ref{ap10}) is increasing in $r$ we arrive at (\ref{root2}).
\qed

One can see that for  $\delta\rightarrow 0$ (less uncertainty), (\ref{ap10}) becomes
\[
\begin{array}{lll}
\ds\frac{r(\sigma^2+\sum_{i=2}^np_i\bar{z}_i)-p_1\bar{z}_1}{\bar{z}_1p_1}\leq 0
&\Leftrightarrow&r(\sigma^2+\sum_{i=2}^np_i\bar{z}_i)-p_1\bar{z}_1\leq 0\\
&\Leftrightarrow&r \leq \ds\frac{p_1\bar{z}_1}{\sigma^2+\sum_{i=2}^np_i\bar{z}_i},
\end{array}
\]
so $r_p$ is the standard ratio
\[
\ds\frac{p_1\bar{z}_1}{\sigma^2+\sum_{i=2}^np_i\bar{z}_i}.
\]
\section*{Appendix II: fundamental inequalities}
\begin{mylem}\label{baslem}It is true that
\begin{equation}\label{lem1}
\ln(1+1/t)\geq 1/(t+1) \quad\forall\ t>0
\end{equation}
\end{mylem}
\Prf  One can easily  check $(t+1)\ln(1+1/t)\geq 1\ \forall\ t>0$ by
plotting the graph of function $(t+1)\ln(1+1/t)$ over $(0,+\infty)$.\qed

\begin{myth}\label{fundth} The function  $f(x,y,t)\triangleq \ln(1+1/xy)^{1/t}$ is convex in the domain
$\{ x>0, y>0, t>0\}$.
\end{myth}
\Prf  Writing $f(x,y,t)=(1/t)(\ln(xy+1)-\ln x-\ln y)$, it is ease to see that the Hessian $\nabla^2 f(x,y,t)$ is
\begin{eqnarray}
\nabla^2 f(x,y,t)&=&
\begin{bmatrix}
\ds\frac{2xy+1}{x^2(xy+1)^2t}&\ds\frac{1}{(xy+1)^2t}&\ds\frac{1}{t^2(xy+1)x}\cr
\ds\frac{1}{(xy+1)^2t}&\ds\frac{2xy+1}{y^2(xy+1)^2t}&\ds\frac{1}{t^2(xy+1)y}\cr
\ds\frac{1}{t^2(xy+1)x}&\ds\frac{1}{t^2(xy+1)y}&\ds\frac{2\ln(1+1/xy)}{t^3}
\end{bmatrix}\nonumber\\
&\succeq&
(x^2y^2(xy+1)^2t^3)^{-1}\begin{bmatrix}
 (xy+1)y^2t^2&x^2y^2t^2&t(xy+1)xy^2\cr
x^2y^2t^2&(xy+1)x^2t^2&t(xy+1)x^2y\cr
t(xy+1)xy^2&t(xy+1)x^2y&2(xy+1)x^2y^2
\end{bmatrix},\label{mat1}
\end{eqnarray}
where the inequality (\ref{lem1}) has been applied to the $(3,3)$-th entry of $\nabla^2 f(x,y,t)$ to
arrive the matrix inequality in (\ref{mat1}).  Here and after,
${\cal A}\succeq {\cal B}$ for matrices ${\cal A}$ and ${\cal B}$ means that ${\cal A}-{\cal B}$ is
a positive definite matrix.
Then, calculating the subdeterminants of matrix in the right hand side (RHS) of (\ref{mat1}) yields
$(xy+1)y^2t^2>0$,
\[
\left|\begin{array}{cc}
\ds\frac{1}{x^2(xy+1)t}&\ds\frac{1}{(xy+1)^2t}\cr
\ds\frac{1}{(xy+1)^2t}&\ds\frac{1}{y^2(xy+1)t}
\end{array}\right|=x^2y^2t^4(2xy+1)>0
\]
and
\[
\left|\begin{array}{ccc}
(xy+1)y^2t^2&x^2y^2t^2&t(xy+1)xy^2\cr
x^2y^2t^2&(xy+1)x^2t^2&t(xy+1)x^2y\cr
t(xy+1)xy^2&t(xy+1)x^2y&2(xy+1)x^2y^2
\end{array}\right|=x^4y^4t^4(xy+1)[(xy+1)^3-1]>0.
\]
Therefore the matrix in the RHS of (\ref{mat1}) is positive definite, implying that the Hessian $\nabla^2 f(x,y,t)$
is positive definite, which is the necessary and sufficient condition for the convexity of $f$ \cite{Tuybook}.\qed

As the function $f(x,y)\triangleq \ln(1+1/xy)$ is convex in the domain $\{x>0, y>0\}$ it follows that \cite{Tuybook}
for every $x>0$, $y>0$, $\bar{x}>0$ and $\bar{y}>0$,
\begin{eqnarray}
\ds\ln(1+1/xy)&=&f(x,y)\nonumber\\
&\geq&f(\bar{x},\bar{y})+\la \nabla f(\bar{x},\bar{y}), (x,y)-(\bar{x},\bar{y})\ra\nonumber\\
&=&\ds\ln(1+1/\bar{x}\bar{y})+\ds\frac{1/\bar{x}\bar{y}}{1+1/\bar{x}\bar{y}}(2-x/\bar{x}-
y/\bar{y}).
\label{sec5.2}
\end{eqnarray}
Similarly, for the convex function $f(x,y,t)\triangleq \ln(1+1/xy)^{1/t}$, one has the following
inequality for every $x>0$, $y>0$, $t>0$, $\bar{x}>0$, $\bar{y}>0$ and $\bar{t}>0$,
\begin{eqnarray}
\ds\ds\frac{\ln(1+1/xy)}{t}&=&f(x,y,t)\nonumber\\
&\geq&f(\bar{x},\bar{y},\bar{t})+\la \nabla f(\bar{x},\bar{y},\bar{t}), (x,y,t)-(\bar{x},\bar{y},\bar{t})\ra\nonumber\\
&=&\ds\ds\frac{2\ln(1+1/\bar{x}\bar{y})}{\bar{t}}
+\ds\frac{1/\bar{x}\bar{y}}{\bar{t}(1+1/\bar{x}\bar{y})}(2-x/\bar{x}-y/\bar{y})
-\ds\frac{\ln(1+1/\bar{x}\bar{y})}{\bar{t}^2}t
\label{esec5.2}
\end{eqnarray}
Analogously, the  inequality
\begin{eqnarray}
\ds\frac{-\ln (1+x)}{t}\geq 2\ds\frac{\alpha-\ln(1+\bar{x})}{\bar{t}}+\ds\frac{\bar{x}}{\bar{t}(1+\bar{x})}-\ds\frac{x}{\bar{t}(1+\bar{x})}
-\ds\frac{\alpha-\ln(1+\bar{x})}{\bar{t}^2}t-\ds\frac{\alpha}{t}\nonumber\\
\forall\ 0\leq x\leq M,  \alpha\geq \ln(1+M)+0.5 \label{nin1}
\end{eqnarray}
follows from the convexity of function $\ds\frac{\alpha-\ln(1+x)}{t}$ over the trust region $0\leq x\leq M$.

Lastly, the inequality
\begin{equation}\label{conv1}
\ds\ln (1+x/y)\leq \ln(1+\bar{x}/\bar{y})+\ds\frac{1}{1+\bar{x}/\bar{y}}
(0.5(x^2/\bar{x}+\bar{x})/y-\bar{x}/\bar{y})
\end{equation}
follows from the concavity of function $\ln(1+z)$ and then the inequality
\begin{eqnarray}\label{conv3}
x&=&0.5(x^2/\bar{x}+\bar{x})-0.5(x-\bar{x})^2/\bar{x}\nonumber\\
&\leq& 0.5(x^2/\bar{x}+\bar{x}) \quad\forall x>0, \bar{x}>0.
\end{eqnarray}
\vspace*{-1.0cm}
\bibliographystyle{ieeetr}
\bibliography{MCMA_Bib}

\end{document}